\documentclass[11pt]{emulateapj}
\usepackage{natbib} 
\usepackage{amsmath}
\usepackage{hyperref}

\begin{document}
\title{IN-SYNC III: The dynamical state of IC 348 - A super-virial velocity dispersion and a puzzling sign of convergence}

\author{Michiel Cottaar\altaffilmark{1}, \and Kevin R. Covey\altaffilmark{2,3}, \and Jonathan B. Foster\altaffilmark{4}, \and Michael R. Meyer\altaffilmark{1}, \and Jonathan C. Tan\altaffilmark{5}, \and David L. Nidever\altaffilmark{6}, \and S. Drew Chojnowski\altaffilmark{7}, \and Nicola da Rio\altaffilmark{5}, \and Kevin M. Flaherty\altaffilmark{8}, \and Peter M. Frinchaboy\altaffilmark{9}, \and Steve Majewski\altaffilmark{7}, \and Michael F. Skrutskie\altaffilmark{7}, \and John C. Wilson\altaffilmark{7} \and Gail Zasowski\altaffilmark{7,10,11}}
\email{MichielCottaar@gmail.com}

\altaffiltext{1}{Institute for Astronomy, ETH Zurich, Wolfgang-Pauli-Strasse 27, 8093 Zurich, Switzerland}
\altaffiltext{2}{Lowell Observatory, Flagstaff, AZ 86001, USA}
\altaffiltext{3}{Department of Physics and Astronomy, Western Washington University, 516 High Street, Bellingham WA 98225 USA}
\altaffiltext{4}{Yale Center for Astronomy and Astrophysics, Yale University New Haven, CT 06520, USA}
\altaffiltext{5}{Department of Astronomy, University of Florida, Gainesville, FL 32611, USA}
\altaffiltext{6}{Department of Astronomy, University of Michigan, Ann Arbor, MI 48109, USA}
\altaffiltext{7}{Department of Astronomy, University of Virginia, Charlottesville, VA 22904, USA}
\altaffiltext{8}{Astronomy Department, Wesleyan University, Middletown, CT, 06459, USA}
\altaffiltext{9}{Department of Physics \& Astronomy, Texas Christian University, Fort Worth, TX 76129, USA}
\altaffiltext{10}{Department of Astronomy, The Ohio State University, Columbus, OH 43210, USA}
\altaffiltext{11}{Center for Cosmology and Astro-Particle Physics, The Ohio State University, Columbus, OH 43210, USA}
\begin{abstract}
Most field stars will have encountered the highest stellar density and hence the largest number of interactions in their birth environment. Yet the stellar dynamics during this crucial phase are poorly understood. Here we analyze the radial velocities measured for 152 out of 380 observed stars in the 2-6 Myr old star cluster IC 348 as part of the SDSS-III APOGEE. The radial velocity distribution of these stars is fitted with one or two Gaussians, convolved with the measurement uncertainties including binary orbital motions. Including a second Gaussian improves the fit; the high-velocity outliers that are best fit by this second component may either (1) be contaminants from the nearby Perseus OB2 association, (2) be a halo of ejected or dispersing stars from IC 348, or (3) reflect that IC 348 has not relaxed to a Gaussian velocity distribution. We measure a velocity dispersion for IC 348 of $0.72 \pm 0.07$ km s$^{-1}$ (or $0.64 \pm 0.08$ km s$^{-1}$ if two Gaussians are fitted), which implies a supervirial state, unless the gas contributes more to the gravitational potential than expected. No evidence is found for a dependence of this velocity dispersion on distance from the cluster center or stellar mass. We also find that stars with lower extinction (in the front of the cloud) tend to be redshifted compared with stars with somewhat higher extinction (towards the back of the cloud). This data suggests that the stars in IC 348 are converging along the line of sight. We show that this correlation between radial velocity and extinction is unlikely to be spuriously caused by the small cluster rotation of $0.024 \pm 0.013$ km s$^{-1}$ arcmin$^{-1}$ or by correlations between the radial velocities of neighboring stars. This signature, if confirmed, will be the first detection of line-of-sight convergence in a star cluster. Possible scenarios for reconciling this convergence with IC 348's observed supervirial state include: a) the cluster is fluctuating around a new virial equilibrium after a recent disruption due to gas expulsion or a merger event, or b) the population we identify as IC 348 results from the chance alignment of two sub-clusters converging along the line of sight. Additional measurements of tangential and radial velocities in IC 348 will be important for clarifying the dynamics of this region, and informing models of the formation and evolution of star clusters. The radial velocities analyzed in this paper have been made available online.
\end{abstract}

\maketitle
\section{Introduction}
\label{sec-1}
\label{sec:intro}
The early (i.e., few Myr) dynamical evolution of star clusters is still poorly constrained. Prestellar and protostellar cores have a small velocity dispersion of 300 - 500 m s\(^{-1}\) \citep{Andre07, Kirk07}, typically smaller than the dispersion measured from the linewidths of low-density gas tracers. \citet{Kirk07} suggested that this dispersion is smaller than the dispersion needed to prevent collapse. This subvirial initial state should cause stars formed from the prestellar cores to fall into the potential well of the molecular cloud after they decouple from any support of the surrounding gas (e.g. from the magnetic fields). Indeed young stars typically appear to have a larger velocity dispersion of 1-2 km s\(^{-1}\) \citep{Joergens06, Furesz06, Furesz08, Tobin09}, however these stellar velocities have not been measured in the same regions as the velocities of the pre-stellar cores, precluding a direct comparison. In a companion paper \citep{Foster15} analyze the radial velocity distribution of NGC 1333 to show that the stellar velocity dispersion is larger than the velocity dispersion of the pre-stellar cores in the same region and hence provide evidence for an initial global collapse (although there are alternative scenarios to explain the increase in the velocity dispersion, such as the increase of the velocity dispersion from the dynamic ejections from unstable multiple systems).

If such an initial collapse takes place it should only last for about a free-fall time. Afterwards the dynamical evolution of the cluster will be dominated by mass loss, both stellar mass loss from for example stellar winds and gas mass loss from the dissipation of the natal molecular cloud. The latter is referred to as gas expulsion and is traditionally thought to play a crucial role in the evolution of a young embedded cluster, because in most star-forming regions up to \(\sim 10\%\) of the molecular gas gets converted into stars (i.e., overall star-formation efficiency; \citealt{Evans09}). Simulations and analytical approximations have shown that if the stars have the same spatial distribution as the gas and are in virial equilibrium the star-formation efficiency will have to be greater than 20-40\% to survive gas expulsion \citep{Tutukov78, Hills80, Elmegreen83, Lada84, Goodwin06}. So local clusters are unlikely to survive gas expulsion, unless either (1) their star-formation efficiency is much higher than average, (2) they are still subvirial at the time of gas expulsion \citep{Goodwin09}, (3) or the stellar group has contracted into the center of the molecular cloud, creating locally a higher effective star-formation efficiency \citep{Smith11, Kruijssen12}. 

To observationally constrain the dynamical evolution during this crucial epoch, we need to observe stellar velocities in clusters before and shortly after this gas expulsion. Here we analyze stellar radial velocities derived as part of the INfrared Spectra of Young Nebulous Clusters (IN-SYNC) ancillary program of the Apache Point Observatory Galactic Evolution Experiment (APOGEE; \citealt{Zasowski13} and Majewski et al. in prep.) from the third Sloan Digital Sky Survey (SDSS-III; \citealt{Eisenstein11}). \citet{Foster15} analyzed the velocity distribution in NGC 1333 and Da Rio et al. (in prep) will look at the velocity distribution in Orion A. Here we focus on IC 348, a young star-forming region in the Perseus molecular cloud \citep[see review from][]{Herbst08}. Age estimates for the stars in IC 348 vary from 2-3 Myr \citep[e.g.,][]{Luhman03} to \(\sim 6\) Myr \citep{Bell13a}. For the velocity dispersion and half-mass radius presented in this work we find that the stars cross the half-mass radius in 0.7 Myr, so for both age estimates IC 348 is many crossing times old. This implies that any initial spatial substructure should have dissipated \citep{Parker14}. Indeed no evidence for spatial substructure in the stellar distribution was found in IC 348 \citep{Muench07, Kumar07, Schmeja08}. Any contraction due to the initial subvirial state described above should also have ended on the free-fall timescale of about 1 Myr and the subsequent evolution should be dominated by mass loss from winds and perhaps supernovae (although no evidence of supernova bubbles have been found; \citealt{Ridge06a}) and the dissipation of the molecular cloud out of which IC 348 formed. This implies that IC 348 should either be in virial equilibrium or should be expanding due to recent mass loss. In this paper we will show that the velocity dispersion of IC 348 does indeed suggest that the cluster is supervirial. However, we will also show that the stars in IC 348 are actually converging along the line of sight, despite its observed virial state and the theoretical arguments above.

In Section \ref{sec:obs} we will discuss the selection of the 380 observed stars, as well as the subset of 152 stars for which we analyze the radial velocity distribution. Here we also summarize the observations and the derivation of the stellar parameters including the radial velocity from high-resolution H-band APOGEE spectra. The spectral analysis is described in detail in \citet{Cottaar14a}. In Section \ref{sec:method} we discuss how we analyze the observed velocity distribution in order to derive the mean velocity and velocity dispersion, corrected for measurement uncertainties and the velocity offsets due to binary orbital motions. The resulting mean velocity and velocity dispersion and their dependence on other stellar parameters such as stellar position, mass, and interstellar extinction are discussed in Section \ref{sec:disp} (for the velocity dispersion) and in Section \ref{sec:mean} (for the mean velocity). In Section \ref{sec:ext} we will present the main result of the paper, namely that stars with larger extinctions are on average blueshifted compared with stars with lower extinction. In Section \ref{sec:disc} we will look at possible scenarios to explain this correlation and conclude that the stars in IC 348 are probably converging along the line of sight. There we also discuss the virial state of IC 348 and interpret a velocity gradient observed across the plane of the sky. Finally our conclusions are presented in Section \ref{sec:conc}.
\section{Observations}
\label{sec-2}
\label{sec:obs}
\subsection{Target selection}
\label{sec-2-1}
\label{sec:target}
Here we briefly discuss how the targets for the APOGEE/IN-SYNC spectroscopic survey of IC 348 were selected. We first describe the creation of the target catalogue and analyze its completeness. Then we discuss the completeness of the subset of the stars in the catalogue for which spectra were actually taken.

Likely IC 348/Perseus members were selected from a catalog of confirmed and candidate IC 348 members compiled by August Muench (private communication), supplemented with mid-infrared excess sources identified by the cores-to-discs (c2d) Spitzer survey team \citep{Jorgensen06, Rebull07}.  The Muench catalog includes 449 confirmed and candidate IC 348 members; the majority of these sources were drawn from previously published catalogs of photometrically selected, and often spectroscopically confirmed, members \citep[e.g.][]{Luhman98, Luhman99, Luhman03, Luhman05, Muench03, Muench07}. Including candidates selected on the basis of deep X-ray observations of the cluster center \citep[i.e.][]{Preibisch01, Preibisch03}, this catalog is likely highly complete in the cluster center \citep[more than 80\% complete for $H < 16$ according to][]{Muench07}, as verified by our independent analysis of source counts in the region (see Figure 2, below). In the more poorly studied outskirts of IC 348, the Muench catalog does include additional candidates selected via an R vs. R-J color-magnitude cut performed utilizing photometry from the USNO NOMAD catalog; these candidates are less secure, particularly at early types, where member selection via the R vs. R-J CMD becomes less efficient due to the steep cluster sequence, and further from the cluster core, where source contamination from background stars and members of the Perseus OB2 association is expected to become significant.

\begin{figure}[htb]
\centering
\includegraphics[width=.9\linewidth]{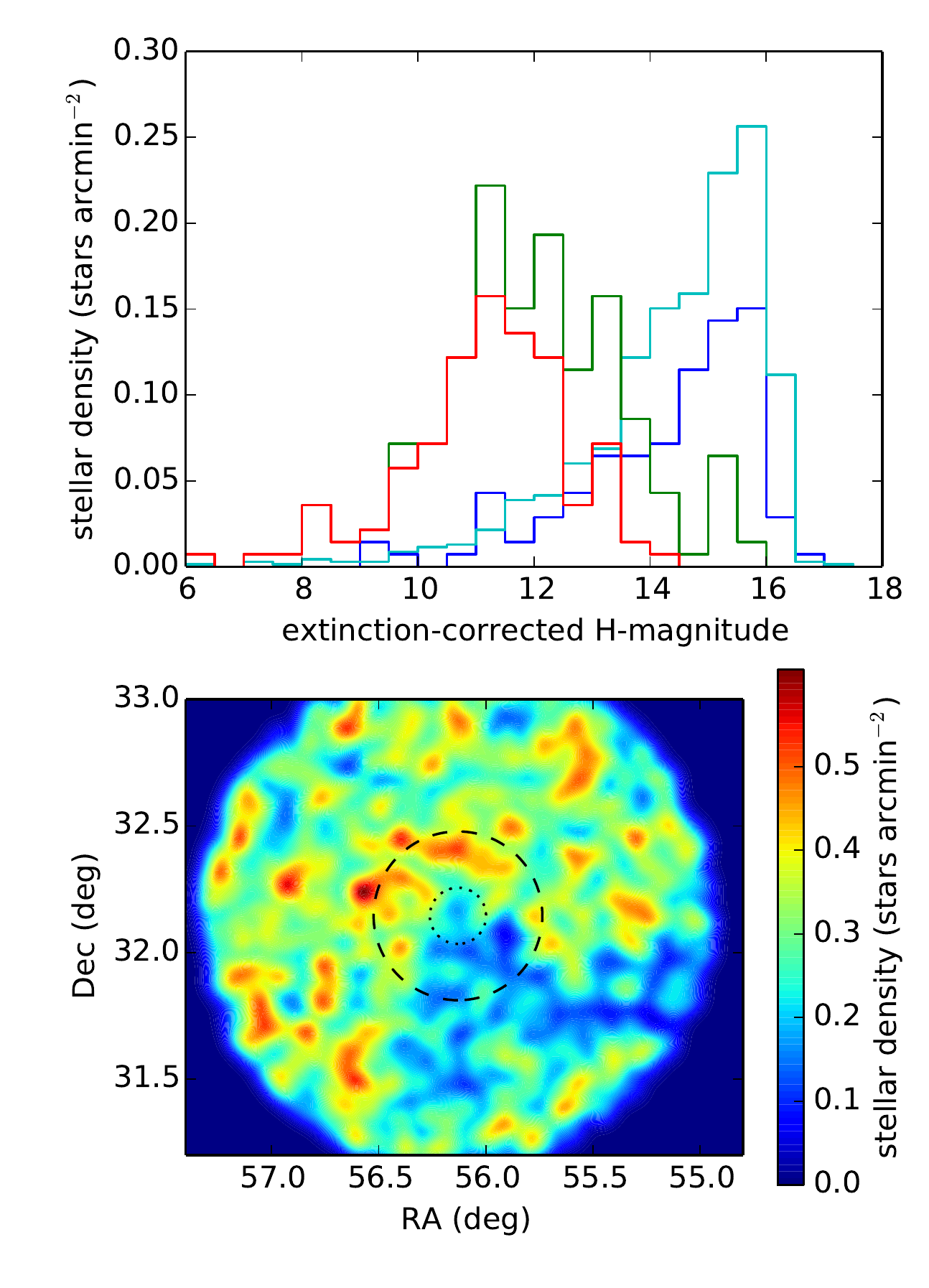}
\caption{\label{fig:complete}The upper panel shows the stellar density distribution per bin of H-band, corrected for the background extinction as estimated from the submillimeter continuum observed by Herschel. The various histograms show the stellar density of the stars in the target catalogue (green), the stars actually observed (red), the 2MASS sources classified as background within the cluster center (within 6.6 arcminutes; blue), and 2MASS sources outside of the cluster (further than 20'; cyan). The bottom panels shows the density distribution of all 2MASS sources not classified as potential members with H\(<13.5\). The two circles are at 6.6 (dotted) and 20 (dashed) arcminutes from the cluster center (same as in Figure \ref{fig:target}). The lack of an overdensity of 2MASS sources classified as background at the position of IC 348 provides evidence for the completeness of the adopted targeting catalog.}
\end{figure}

We provide evidence for the completeness of the target catalogue in the cluster core by studying the distribution of the 2MASS sources that are not in our target catalogue in a similar manner as \citet{Cambresy06}. If the target catalogue is complete, all stars not in the catalogue should be background and hence should have the same brightness distribution and spatial density as background stars in an off-cluster field (after correcting for extinction). The upper panel in Figure \ref{fig:complete} shows the extinction-corrected H-band density distribution of these untargeted 2MASS sources in the cluster core (blue) and the 2MASS sources in an off-cluster field covering 20' to 60' from the cluster center (cyan). In the magnitude range covered by our spectroscopic observations (\(H < 13.5\)), the distributions are consistent, suggesting that the targeting catalogue is (mostly) complete. This is confirmed in the spatial distribution of non-targeted 2MASS sources with \(H < 13.5\) (lower panel of Figure \ref{fig:complete}), which shows no remaining overdensity of sources at the location of IC 348.

\begin{figure}[htb]
\centering
\includegraphics[width=.9\linewidth]{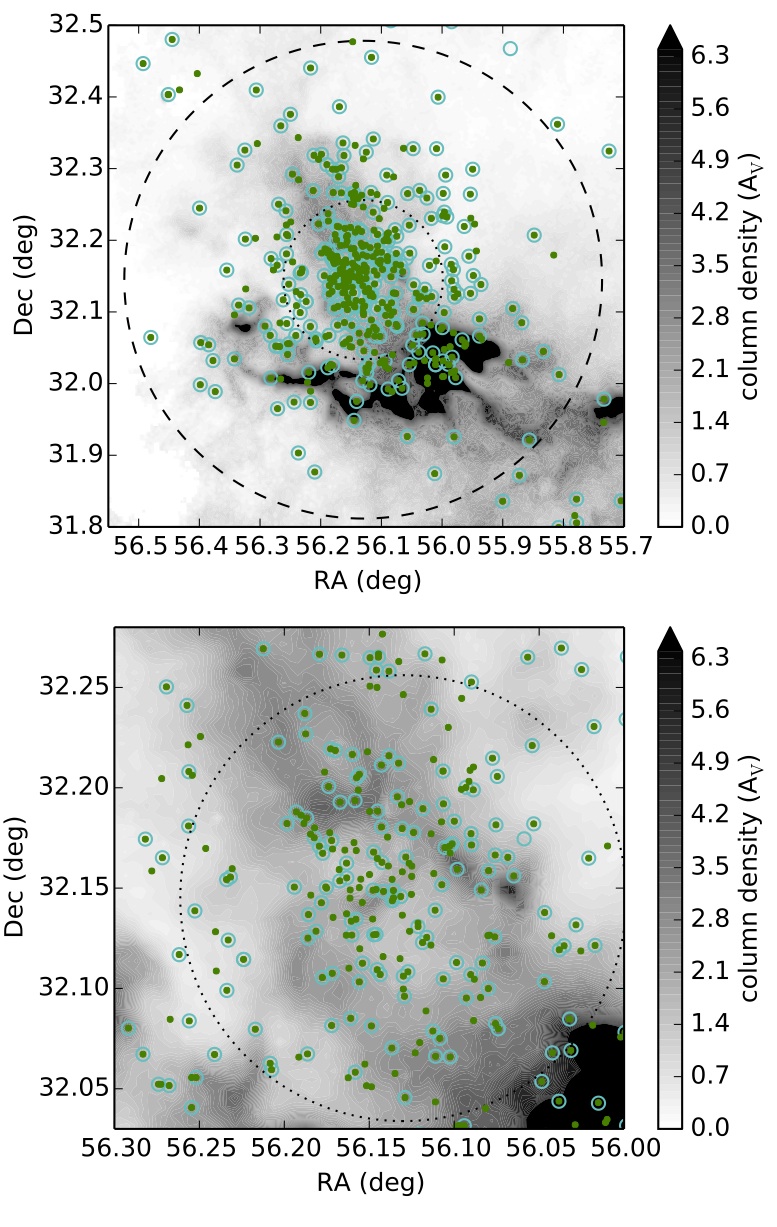}
\caption{\label{fig:target}Map of the stars in the target catalogue in green. Stars circled in cyan have been observed. Additional background stars observed to fill up the APOGEE fibers have not been marked. The lower panel shows a zoom-in of the center of the upper panel, illustrating the lower coverage in the cluster center. For comparison with Figure \ref{fig:complete} we added two circles at 6.6 (dotted) and 20 (dashed) arcminutes from the cluster center. The background shows a Herschel column density map (see Section \ref{sec:gas}).}
\end{figure}

Although the target catalogue appears to be reasonably complete (at least in the cluster core), the spectroscopically observed stars are not complete. Most targets in the outskirts were observed, however only about half of the stars in the target catalogue in the cluster core were observed (Figure \ref{fig:target}), because APOGEE is unable to simultaneously target stars within \(\sim 71.5\)" of each other due to a collision through fiber assignment. This meant that even though 9 separate plates were drilled to cover IC 348, many stars in the dense cluster core could still not be targeted. 

The priority of the targets was based on the H-band magnitude. The highest priority was given to targets with \(7 < H < 12.5\). For these high priority target we are mostly complete (\(\sim 90 \%\)), even in the cluster center (upper panel in Figure \ref{fig:complete}). For fainter targets the priority was assigned based on the H-band magnitude after correction of the background extinction, where the extinction was estimated from the J-H vs. H-Ks 2MASS color-color diagram. Intrinsically brighter stars were targeted first.

Most (90 \%) stars were observed for multiple epochs. Half of these were observed within a single year (with a baseline up to a few months), while the other half were observed over two years with a baseline up to 500 days. The spectra have a median signal to noise of 70.

\subsection{Spectral analysis}
\label{sec-2-2}
\label{sec:spec}
\citet{Cottaar14a} describes the analysis of the high-resolution near-infrared spectra obtained by the APOGEE multi-object spectrograph \citep{Wilson12} from stars in IC 348, NGC 1333, NGC 2264, and Orion A as part of the IN-SYNC ancillary program. In summary BT-Settl model spectra \citep{Allard11} were fitted to every reduced spectrum \citep{Nideversubm} in a minimum chi-squared sense after masking any bad pixels or strong telluric emission lines. From our best-fit we extract the following parameters: the effective temperature, surface gravity, H-band veiling, rotational velocity, and radial velocity. During these fits the observed continuum was matched to the continua of the model spectra using a polynomial. Accurate estimates of the interstellar reddening were obtained by computing the E(J-H) of the stars in IC 348 with respect to an empirical color locus in the Pleiades, which we mapped to our effective temperature scale by measuring effective temperatures from APOGEE spectra observed for stars in the Pleiades using the same pipeline as for IC 348. \citet{Cottaar14a} showed for a subset of the stars in IC 348 for which optical photometry was available that the E(J-H) measured in the infrared accurately predicted the reddening observed in the optical, which put an upper limit on the reddening uncertainty of only about 0.06 in E(J-H).

Throughout this paper we will focus on the observed radial velocities. \citet{Cottaar14a} found a systematic redshift for the coolest stars in IC 348 (as well as NGC 1333) of a few km s\(^{-1}\), which we argued was likely caused by inaccuracies in the molecular (probably water) line lists affecting stars cooler than 3300 K. An empirical spline was fitted to this systematic offset and subtracted to get a consistent zero-point across all stellar masses. This corrected radial velocity will be used throughout this paper. The radial velocities were not corrected for variations of the velocity zero-point over time, which were found to be small (up to about 100 m s${-1}$) in the APOGEE survey \citep{Nideversubm}. The current APOGEE pipeline estimates typical radial velocity uncertainties for our sample of around 100 m s$^{-1}$. In our analysis, we scale those uncertainties to match the sample's measured epoch-to-epoch RV variability, which is larger by a factor of $\sim$3 than the APOGEE pipeline's estimates (see discussion of this effect by \citep{Cottaar14a} in their Sec. 3.2.1).  As a result, the kinematic results that follow should be robust against all random uncertainties. The radial velocities (Table \ref{tab:vel}) have been made available online with this publication. The other parameter have been previously released by \citet{Cottaar14a}.

\begin{table*}
\caption{\label{tab:vel}Sample of the radial velocities of stars in IC 348. The fourth column is the radial velocities corrected for the systematic redshift found for the coolest stars \citep{Cottaar14a}. The full table has been made available on the journal website and \url{http://www.astro.ufl.edu/insync/}.}
\centering
\begin{tabular}{|c|c|c|c|c|}
2MASS & Julian date & radial velocity (km/s) & corrected $v_{\rm rad}$ (km/s) & uncertainty v$_{\rm rad}$ (km/s) \\
\hline
2M03233718+3056336 & 2456564.00 & -98.50 & -98.50 & 0.27 \\
2M03233718+3056336 & 2456607.75 & -98.33 & -98.33 & 0.26 \\
2M03233787+3131094 & 2456561.75 & -2.03 & -2.03 & 0.31 \\
2M03233787+3131094 & 2456236.75 & -1.95 & -1.95 & 0.26 \\
2M03233787+3131094 & 2456315.75 & -1.99 & -1.99 & 0.24 \\
2M03234277+3053142 & 2456564.00 & -20.57 & -20.57 & 0.30 \\
2M03234277+3053142 & 2456607.75 & -20.73 & -20.73 & 0.12 \\
2M03234517+3109561 & 2456674.50 & 37.55 & 37.55 & 0.16 \\
2M03234831+3121526 & 2456674.50 & 25.74 & 25.74 & 0.20 \\
2M03235905+3101512 & 2456671.50 & 48.43 & 48.43 & 0.32 \\
2M03235905+3101512 & 2456674.50 & 48.11 & 48.11 & 0.28 \\
2M03235905+3101512 & 2456564.00 & 47.87 & 47.87 & 0.39 \\
2M03235905+3101512 & 2456607.75 & 47.91 & 47.91 & 0.36 \\
2M03240225+3103502 & 2456674.50 & -36.80 & -36.72 & 0.20 \\
\end{tabular}
\end{table*}

\subsection{Sample for velocity analysis}
\label{sec-2-3}
\label{sec:samp}
We make several cuts to the total population of observed cluster members to define a population of IC 348 members useful for a dynamical analysis of the cluster. The main goal of these cuts is to only analyze stars for which the radial velocity has been accurately determined and for which this radial velocity is likely to be affected by the gravitational field of IC 348. First we make two cuts to the individually observed spectra:
\begin{enumerate}
\item From the 2323 observed spectra in IC 348 we cut 591 spectra with S/N less than 20. These spectra are so noisy that they provide very little information on the actual radial velocity of the star. Furthermore the convergence in the spectral fitting is often very poor for these low S/N spectra, which leads to offsets in the stellar parameters larger than expected from the already large noise level.
\item From the remaining 1732 spectra we cut an additional 17 spectra with best-fit effective temperature less than 2400 K. Our survey does not go deep enough to actually observe such cool stars, which suggests that these rare low best-fit effective temperatures represent fits to the noise in the spectra and not fits to the spectral features. Indeed all of the stars from which these spectra were taken have other epochs with more reasonable stellar parameters including larger effective temperatures.
\end{enumerate}

\begin{figure}[htb]
\centering
\includegraphics[width=.9\linewidth]{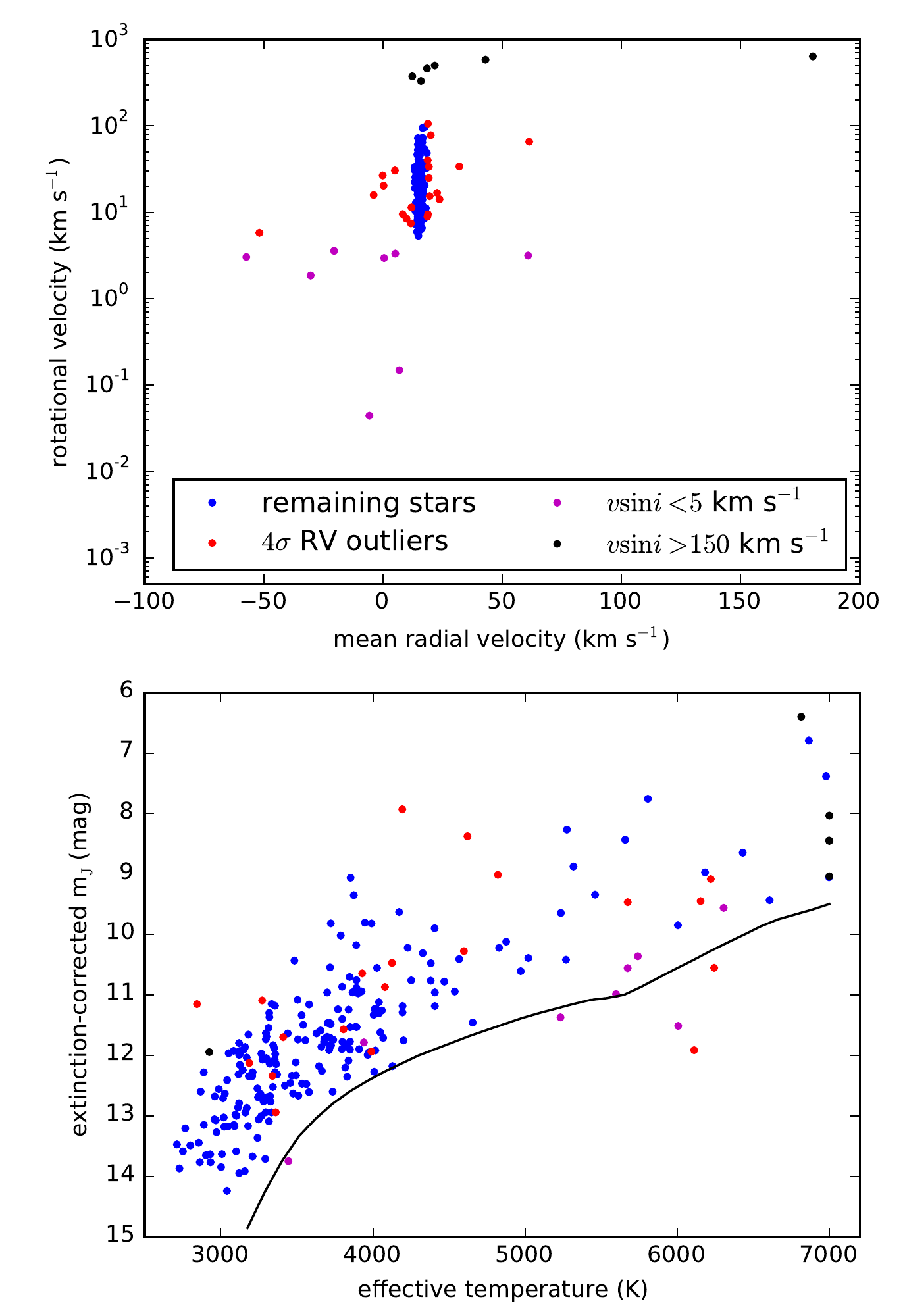}
\caption{\label{fig:clean}Two plots of the stellar parameter distribution for slow rotators (purple; \(v \sin i < 5\) km s\(^{-1}\)) and rapid rotators (black; \(v \sin i > 150\) km s\(^{-1}\)). The remaining population of intermediate rotators is split into those stars with radial velocities within 3 km s\(^{-1}\) (i.e., \(\sim 4\) times the velocity dispersion of IC 348) of the mean velocity (blue) and those with a radial velocity offset from the mean larger than 3 km s\(^{-1}\) (red). The top panel show the weighted mean radial velocity over all epochs versus the rotational velocity \(v \sin i\). The bottom panel shows the effective temperature versus the extinction-corrected J-band magnitude.}
\end{figure}

For 67 out of 380 stars observed in IC 348, no spectra survive the cuts described above. So after these initial cuts we are left with 313 stars. This is further reduced to a total used sample size of 152 by cutting any stars that fall within one or more of the following categories:
\begin{enumerate}
\item 12 stars with \(v \sin i < 5\) km s\(^{-1}\) are cut, because they are very unlikely to be actual cluster members. These stars are plotted in purple in Figure \ref{fig:clean}. They have a velocity distribution with no clear peak around the velocity of IC 348 (top panel) and are typically hot stars (KS-test p-value \(= 10^{-6}\) that they have the same effective temperature distribution as the other stars), which are relatively faint (bottom panel). These stars are likely contaminants, which ended up being observed based on their position in the R vs. R-J CMD.
\item 11 stars with \(v \sin i > 150\) km s\(^{-1}\) are cut. These are likely to be cluster members, however most of these stars appear to be very early type with 8 out of 11 stars having a best-fit effective temperature of 7000 K (black points in Figure \ref{fig:clean}), which is the upper bound in the adopted grid of model spectra. The predominance of the hydrogen lines for these early-type stars might have caused an overestimate of their large rotational velocities, as well as a lower reliability of the radial velocities of these stars as shown by the increased spread in RV for these stars in the upper panel of Figure \ref{fig:clean}. This increased uncertainty in the radial velocities is the main reason to exclude these stars.
\item We cut 63 stars, which are more than 20' ($>$ 3 half-mass radii) from the cluster center, because they are far enough that they are unlikely to be gravitationally affected by the cluster and hence do not provide us information about the current dynamical state of the cluster. Many of these stars show clear signs of youth and they might represent a population from distributed star-formation in the Perseus molecular cloud, they might be related to the Perseus OB2 association in which IC 348 is embedded, or they might be ejected by IC 348. We will briefly discuss these outlying stars in Section \ref{sec:cuteff}.
\item We cut 89 stars with uncertainties on the radial velocity averaged over all epochs larger than 500 m s\(^{-1}\). These stars have velocity uncertainties comparable to the velocity dispersion itself and hence their velocity distribution is not dominated by the intrinsic velocity distribution of IC 348, but by the measurement uncertainties. These measurement uncertainties are themselves uncertain and appear to be non-Gaussian \citep{Cottaar14a}. As we will show in Section \ref{sec:cuteff} the effect of this cut on the fit to the velocity distribution is very small, because the measurement uncertainties are explicitly taken into account in the fitting procedure.
\item Out of the remaining 172 stars after the cuts described above, 20 more are cut because they are RV-variable. To define a star as RV-variable we first compute the \(\chi^2\) for a model where the RV is the same over all epochs: 
\begin{equation}
\chi^2 = \sum_i \frac{({\rm RV}_i - \mu)^2}{\sigma_i^2},
\end{equation}
where we sum the square of the radial velocity offset from the weighted mean (\({\rm RV}_i - \mu\)) normalized by the measurement uncertainty \(\sigma_i\) over all epochs \(i\). If the probability of drawing the computed $\chi^2$ or a larger \(\chi^2\) due to random noise is smaller than \(10^{-4}\) we conclude the star is RV-variable. This probability is computed from a $\chi^2$-distribution with the degrees of freedom set to the number of epochs minus one.
\end{enumerate}

None of these cuts are designed to match the mass or spatial distribution of the adopted sample to the actual mass or spatial distribution of all IC 348 cluster members. This means that if the stellar velocity dispersion depends on stellar mass or location, the velocity dispersion of the sample analyzed here will not be representative of the dynamical state of all stars in IC 348. We will show that these possible biases have a limited effect in section \ref{sec:radial} for the spatial bias and in section \ref{sec:mass} for the bias in stellar mass.
\section{Method to fit the velocity distribution}
\label{sec-3}
\label{sec:method}
The observed radial velocity distribution of our analyzed sample of 152 stars (with a total of 758 observed spectra) in IC 348 is not only influenced by the stellar motions through the cluster's potential, but also by orbital motions of binary members and potential contamination of background or foreground stars, especially from young stars from the surrounding Perseus OB association. All these effects create a complex non-Gaussian velocity distribution. We model this distribution to retrieve the underlying dynamical properties of IC 348 using the maximum-likelihood procedure outlined in \citet{Cottaar12a} and \citet{Cottaar14}. 

In summary we start by assuming that IC 348 has intrinsically a Gaussian velocity distribution:
\begin{equation}
  P(v_{\rm intr}) = \frac{1}{\sqrt{2 \pi \sigma_c^2}} e^{-(v_{\rm intr} - \mu_c)^2 / 2 \sigma_c^2} \label{eq:vintr},
\end{equation}
where \(P(v_{\rm intr})\) is the probability distribution of the intrinsic velocity of the star (\(v_{\rm intr}\)) and \(\mu_c\) and \(\sigma_c\) are respectively the mean velocity and velocity dispersion of the cluster. To actually compare this with the observed velocity distribution we have to convolve this distribution of intrinsic velocities with the probability distribution of the measurement uncertainties (\(P(v_{\rm unc})\)) and the binary orbital motions (\(P(v_{\rm bin})\)):
\begin{align}
  P(v_{\rm obs}) = &(1 - f'_{\rm bin}) P(v_{\rm intr}) \ast P(v_{\rm unc}) \nonumber \\
                   &+ f'_{\rm bin} P(v_{\rm intr}) \ast P(v_{\rm unc}) \ast P(v_{\rm bin}) \label{eq:conv},
\end{align}
where \(f'_{\rm bin}\) is the binary fraction after removing RV-variable stars, which we will define in equation \ref{eq:fbin_corr}. This convolution will have to be computed separately for every star, because the measurement uncertainties and the effects of binary orbital motions differ per star (see below). Because we only analyze the stars for which the radial velocity has been measured with a precision smaller than 500 m s\(^{-1}\) the width of the measurement uncertainties (\(P(v_{\rm unc})\)) is for all stars smaller than that of the intrinsic velocity distribution (\(P(v_{\rm intr})\)), however for completeness we still do include it as a Gaussian distribution, which is only a rough approximation of the true probability distribution due to the measurement uncertainties \citep{Cottaar14a}.

Computing the velocity offsets due to binary orbital motions ($P(v_{\rm bin})$) is more complicated, because the multiple epochs over which APOGEE spectra were taken allows the binaries with short periods (up to several times the observational baseline) to be detected through their radial velocity variations. However, many binaries with orbital velocity amplitudes comparable to or larger than the velocity dispersion of IC 348 remain undetected, broadening the observed radial velocity distribution. We already excluded from the analysis all stars (20 out of 172) with variable radial velocities (i.e. inconsistent with being single at the \(p < 10^{-4}\) level according to a \(\chi^2\)-test). For the remaining stars we compute the distribution of radial velocity offsets due to binary orbital motions by (1) drawing in a Monte-Carlo like fashion a large number of binary orbits, (2) creating for every star artificial RV observations given the observational epochs and measurement uncertainties for that star, (3) discarding those binary orbits that would have been detected in the artificial RV observations by the \(\chi^2\)-test, and (4) computing the distribution of radial velocity offsets between the observed and systematic velocities for the remaining binaries \citep{Cottaar14}. This radial velocity offset distribution is used as \(P(v_{\rm bin})\) in the convolution in equation \ref{eq:conv}. This same population of binaries is also used to estimate the binary fraction of the seemingly single stars \(f'_{\rm bin}\):
\begin{equation}
  f'_{\rm bin} = f_{\rm bin} \frac{1 - f_{\rm det}}{1 - f_{\rm bin} f_{\rm det}}, \label{eq:fbin_corr}
\end{equation}
where \(f_{\rm bin}\) is the binary fraction for the primary mass of the star prior to any multi-epoch observations and \(f_{\rm det}\) is the fraction of potential binary orbits, which could have been detected in the multi-epoch observations (i.e., the fraction of simulated binary orbits discarded in step 3).

The first step of this procedure requires the drawing of many random binary orbits and hence we need to assume a period, mass ratio, and eccentricity distribution for the observed stars. We are mainly interested in the number of binaries with periods between tens and thousands of years, whose binary orbital motions cause velocity offsets on the order of the velocity dispersion of IC 348. The binary distribution in this period range has only been extensively characterized in the solar neighborhood \citep[see review from][]{Duchene13}, so we will use these results to inform our assumed orbital parameter distributions. First we assign to every star an age and a mass based on the Dartmouth tracks on their effective temperature and extinction-corrected absolute J-band magnitude (assuming a distance modulus of 6.98; \citealt{Ripepi14}). The binary fraction and the mean and width of the semi-major axis distribution all decrease towards lower masses \citep[e.g.][]{Lada06, Janson12}. For brown dwarves we set the binary fraction to 22\% \citep{Burgasser07}. Above the hydrogen burning limit we let the binary fraction increase linearly, so that the binary fraction at one solar mass is 44\% \citep{Raghavan10}. We implement the semi-major axis distribution as three log-normals, with a mean of \(\log a_0 = 0.86\) and a width of \(\sigma_{\log a} = 0.24\) below 0.2 M\(_{\odot}\) \citep{Burgasser07}, with \(\log a_0 = 1.2\) and \(\sigma_{\log a} = 0.8\) between 0.2 M\(_{\odot}\) and 0.8 M\(_{\odot}\) \citep{Janson12}, and with \(\log a_0 = 1.64\) and \(\sigma_{\log a} = 1.52\) above 0.8 M\(_\odot\) \citep{Raghavan10}. We adopt a flat mass ratio distribution \citep{Reggiani13} and flat eccentricity distribution \citep{Duchene13} over the whole mass range. These assumptions about the binary properties based on the field stars are expected to only approximate the true binary properties in IC 348, however \citet{Cottaar12a} showed that the best-fit mean velocity and velocity dispersion are not very sensitive to the adopted binary properties.

The efficiency with which binaries could be detected (as estimated by \(f_{\rm det}\)) varies widely across the sample. Out of the 152 stars whose velocities are analyzed in this paper (i.e. after the cuts described in section \ref{sec:samp}) 20 have only been observed for a single epoch, 33 for two epochs, while the remaining 99 stars have been observed for three up to thirteen epochs. For the stars with multiple epochs, the total observational baseline varies with all observations for 75 stars taken within a single season, while for the remaining 57 stars we have data taken over two years. Finally the RV precision of the observations also varied between 60 and 500 m s\(^{-1}\) with a median of 200 m s\({-1}\). The wide variety of binary orbital parameter distributions assumed for different mass ranges and the variety of observational constraints lead to a binary detection rate (\(f_{\rm det}\)) that ranges from 0 for stars with a single epoch and a few percent for stars below 0.2 M\(_{\odot}\) up to \(\sim 20\%\) for solar-type stars with a baseline of at least a year. However, even in this best-case scenario (i.e. solar mass stars with a baseline of $\sim 1$ year) we can not detect binaries with orbital velocities comparable to the velocity dispersion which would have periods of about $10^5$ years. As a result, despite this careful analysis of these multi-epoch observations, we still expect the majority of relevant binaries to go undetected, so that changes in our binary detection efficiency will have only a limited effect except in the wings of the measured velocity distribution \citep{Cottaar14}.

This procedure allows us to compute the likelihood of observing a given velocity (\(P(v_{\rm obs})\)) from the convolution in equation \ref{eq:conv} given a set of variables parameterizing the intrinsic velocity distribution ($P(v_{\rm intr})$). We consider the best-fit parameters to be those where the total likelihood of reproducing all observed velocities (\(v_{\rm obs}\)) is maximized (i.e., maximize \(\mathcal{L} = \prod P_{\rm obs}(v_{\rm obs})\)). The uncertainties on these parameters will be computed by Markov chain Monte Carlo (MCMC) simulations throughout this paper. To compare the relative accuracy of various models of the mean velocity and velocity dispersion profile we will use the Bayesian Information Criterion (BIC; \citealt{Schwarz78}). This criterion states that the best model is that, which minimizes
\begin{equation}
  \text{BIC} = - 2 \ln \mathcal{L} + k \ln(n), \label{eq:bic}
\end{equation}
where \(\mathcal{L}\) is the best-fit likelihood of reproducing all observed velocities, \(k\) is the number of free parameters, and \(n\) is the number of fitted datapoints (i.e., 152 radial velocities). The first term decreases if the fit is improved, while the second term acts as a penalty on the number of parameters, which prevents overfitting. The BIC values for the models discussed in this paper are shown in Table \ref{tab:BIC}.

Through the paper we vary on this general scheme in two important ways: we either take the mean velocity (\(\mu_c\)) and the velocity dispersion (\(\sigma_c\)) of IC 348 as global variables that are constant across IC 348 or we let them vary as a function of stellar parameters (e.g., stellar mass or location). The rows in Table \ref{tab:BIC} represent different models for the mean velocity and/or velocity dispersion, which we shall discuss throughout the rest of this paper. The second important variation is whether we consider the possibility that a single Gaussian distribution might be insufficient to represent the true intrinsic velocity distribution. To explore this we also model the intrinsic velocity distribution as a sum of two Gaussian, one of which has the IC 348 mean velocity (\(\mu_c\)) and velocity dispersion (\(\sigma_c\)) and the other of which has an alternative mean velocity (\(\mu_a\)) and velocity dispersion (\(\sigma_a\)) \citep[e.g.,][]{Jeffries14}. Such a model has three more free parameters, namely the alternative mean velocity and velocity dispersion and the fraction of analyzed stars, whose velocity distribution is described by this second Gaussian. The columns in Table \ref{tab:BIC} show the result for both a single-Gaussian model and a double-Gaussian model. 

\begin{table*}[htb]
\caption{\label{tab:BIC}Bayesian information criterion (BIC; eq. \ref{eq:bic}) for the fits to the velocity distribution discussed throughout this paper.}
\centering
\begin{tabular}{|c|c|c|c|c|}
Description of fit & Mean velocity & Velocity dispersion & BIC (\# parameters) & BIC (\# parameters)\\
 &  &  & Single Gaussian & Double Gaussian\\
\hline
Global fit & constant & constant & 648 (2) & 523 (5)\\
Polynomial radial dispersion profile & constant & equation \ref{eq:rad_pol} & 658 (4) & 532 (7)\\
Polynomial dispersion profile with mass & constant & equation \ref{eq:mass_pol} & 657 (4) & 531 (7)\\
Step function at \(\sim 0.44\) M\(_{\odot}\) & constant & equation \ref{eq:step_mass} & 656 (4) & 529 (7)\\
Step function at \(\sim 0.8\) M\(_{\odot}\) & constant & equation \ref{eq:step_mass} & 659 (4) & 542 (7)\\
Velocity gradient with extinction & equation \ref{eq:ext} & constant & 636 (5) & 509 (8)\\
Velocity gradient on the sky & equation \ref{eq:rot} & constant & 658 (4) & 532 (7)\\
\end{tabular}
\end{table*}
\section{Trends in velocity dispersion}
\label{sec-4}
\label{sec:disp}
\subsection{Global velocity dispersion}
\label{sec-4-1}
\label{sec:global}
\begin{figure*}
\centering
\includegraphics[width=1\textwidth]{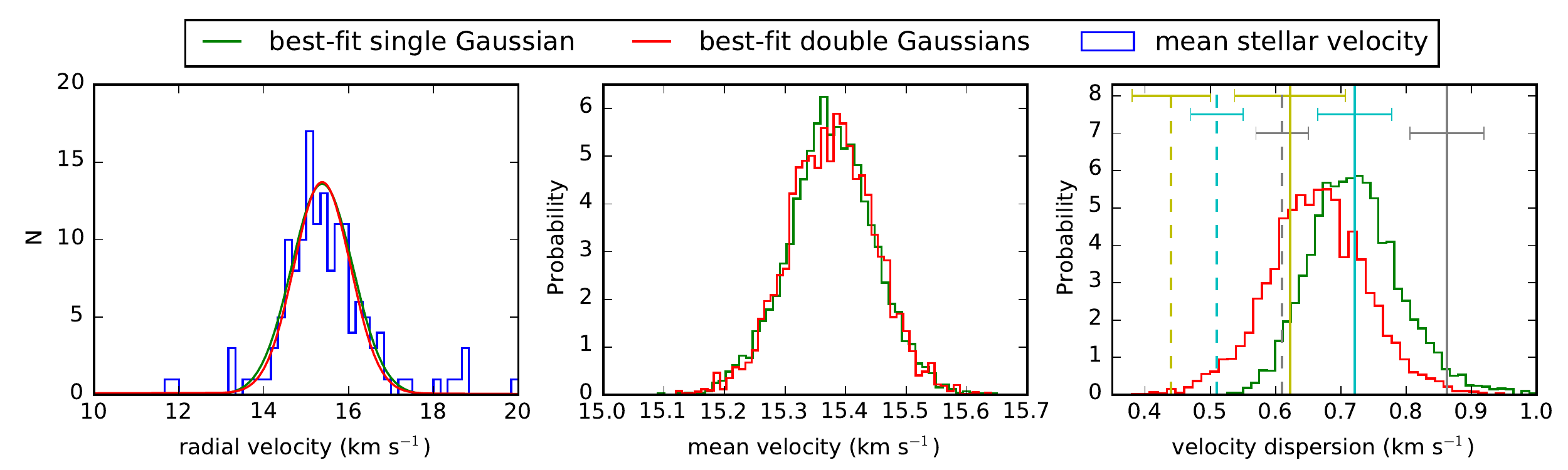}
\caption{\label{fig:vdisp}Left panel: Histogram of the average velocities of all stars included in the fit (blue) with the best-fit intrinsic velocity distribution from the single-Gaussian model (green) and the double-Gaussian model (red). The effect of binary orbital motions on the best-fit models has been included in the fits, but not in the plot, because this effect is different for every star depending on the precision of the velocity measurements and the baseline over which they were observed. Other panels: The probability distribution in the mean velocity and velocity dispersion for the single-Gaussian (green) and the main peak of the double-Gaussian (red) model. The vertical colored lines in the right panel mark the velocity dispersion expected for virial equilibrium (dashed colored lines) and a dynamically unbound state with a net positive energy (towards the right of the solid colored lines) under various assumptions for the contribution of the gas mass to the dynamical state of IC 348, namely yellow for no contribution from the gas, cyan for a small contribution of \(80 M_\odot\), and gray for a large contribution of \(210 M_\odot\) (see Section \ref{sec:virial}).}
\end{figure*}

We start with our simplest model fit to the observed velocity distribution, namely an intrinsic Gaussian velocity distribution with a constant mean velocity and velocity dispersion throughout the cluster, which is convolved with both the measurement uncertainties and the binary orbital motions as described in section \ref{sec:method}. This model only has two free parameters, namely the mean velocity and velocity dispersion of IC 348. The resulting best-fit intrinsic Gaussian velocity distribution has been plotted in green in Figure \ref{fig:vdisp}. The posterior probability distribution of the mean velocity and velocity dispersion resulting from the MCMC simulations have been shown in green in the other two panels. Marginalized over the other parameters the MCMC simulation gives a mean heliocentric velocity of \(15.37 \pm 0.07\) km s\(^{-1}\) \footnote{The uncertainty in the mean velocity is actually dominated by a possible systematic offset with respect to RV standards on the order of about 0.5 km s\(^{-1}\) \citep{Cottaar14a}, not by the statistical uncertainty quoted here.} and a velocity dispersion \(0.72 \pm 0.06\) km s\(^{-1}\). 

The observed velocity distribution does not appear to be truly Gaussian, however, due to an excess of stars at high and low velocities relative to the cluster's mean velocity, compared to the number expected from a Gaussian distribution even after taking into account binary orbital motions. Although visually the small number of "outliers" has a minor effect (left panel in Figure 4), the likelihood of reproducing all observed velocities increases significantly when the fit allows for a profile with stronger high velocity wings.  This is true for tests of profiles with intrinsically stronger wings (i.e., a Lorentzian profile), but the high velocity outliers in our dataset are best modeled with a separate Gaussian component, whose amplitude can be tuned to the number of high velocity objects. Adding this second Gaussian leads to a significant decrease of 125 in the BIC (upper row in Table \ref{tab:BIC}) compared with the single-Gaussian model, which is much larger than 10, the limit usually adopted as providing decisive evidence in favor of a model \citep{Kass95}. This second Gaussian is found to have a mean velocity of \(14 \pm 4\) km s\(^{-1}\) and a velocity dispersion of \(5 \pm 4\) km s\(^{-1}\) with only \(10 \pm 5\%\) of the total stars contributing to this second Gaussian. The large parameter uncertainties are due to the small contribution of this Gaussian and due to its large overlap with the velocity distribution of the first Gaussian. Adding this second Gaussian reduces the velocity dispersion of the main Gaussian to about \(0.64 \pm 0.08\) km s\(^{-1}\) with no significant change in the mean velocity (Figure \ref{fig:vdisp}). Although the double-Gaussian model does provide a significantly better fit than the single-Gaussian model (according to the BIC), we will in this paper continue to consider both models, which generally agree very well with each other on the dynamical properties of IC 348.

The physical interpretation of this second Gaussian is unclear. It could possibly represent contamination from the Perseus OB2 association. However, the mean velocities of the two Gaussians are consistent with each other, which suggests that they are a single population. In that scenario it could be that the second Gaussian represents stars ejected from IC 348 or that the velocity distribution of IC 348 is simply non-Gaussian at this young age. We consider the latter option plausible, because we are unaware of any predictions that a turbulent molecular cloud would produce stars with a Gaussian velocity distribution or that a non-Gaussian initial stellar velocity distribution would relax to a Gaussian one within the few Myrs, as required given IC 348's age. Recent expansion or collapse of IC 348 might also have altered the velocity distribution. Even though the velocity distribution of IC 348 appears to be non-Gaussian, we still adopt the Gaussian model, because it gives us two easily interpretable values (i.e. the mean velocity and the average deviation from this velocity). For the single-Gaussian model the velocity dispersion can be interpreted as the average deviation from the mean velocity for all IC 348 stars, while for the double-Gaussian model the velocity dispersion represents the average deviation for the subset of IC 348 stars with similar velocities, while ignoring the relative large contributions of a few stars with outlying radial velocities.

In this section we will continue to discuss the velocity dispersion and what it means for the dynamics of IC 348. In section \ref{sec:mean} we will then focus on the mean velocity and its spatial dependence.
\subsection{Effects of the sample selection}
\label{sec-4-2}
\label{sec:cuteff}
Here we look at the effect of the cuts described in Section \ref{sec:samp} on the fit to the velocity distribution. In particular we will focus on two cuts, that remove a large number of stars: the 89 stars cut because of their large RV measurement uncertainty (\(> 500\) m s\(^{-1}\)) and the 63 stars cut because they lie more than \(20\arcmin\) from the cluster center.

\begin{figure*}
\centering
\includegraphics[width=1\textwidth]{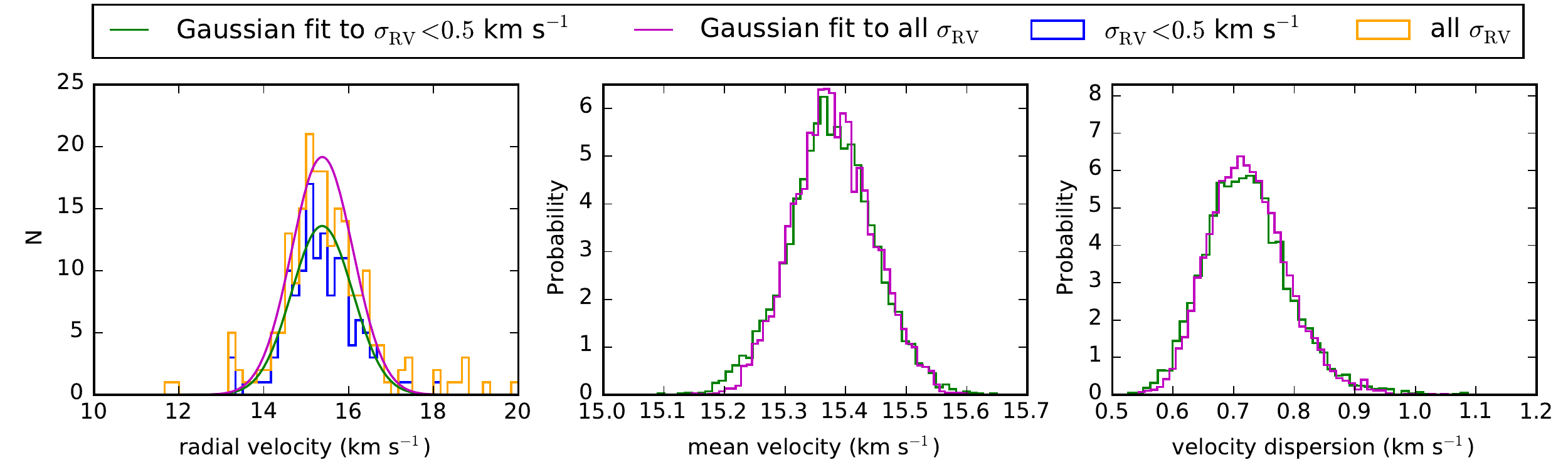}
\caption{\label{fig:bad}Left panel: Histogram of the velocities of all stars included in the fit with the cut in the RV measurement uncertainty at 500 m s\(^{-1}\) (blue) and without this cut (orange). The lines show the best-fit single-Gaussian intrinsic velocity distribution to the radial velocities with the measurement uncertainty cut (green) and without the cut (magenta). Middle panel: Posterior probability distribution of the mean velocity with the cut in the RV measurement uncertainty (green) and without the cut (magenta). Right panel: Same as the middle panel, but with the posterior probability distribution of the velocity dispersion rather than the mean velocity.}
\end{figure*}

Figure \ref{fig:bad} illustrates the changes in the single-Gaussian fit if the stars with large RV measurements are included. Because the RV uncertainties are explicitly included in the fit, the inclusion of the high-uncertainty stars actually has very little effect with shifts of \(\sim 20\) m s\(^{-1}\) in the mean velocity and velocity dispersion and minimal reductions in the precision of these parameters. For the high-uncertainty stars the offset from the mean velocity is mostly set by the measurement uncertainty rather than the velocity dispersion, so they provide little to no information on the velocity dispersion. Similarly, their larger spread implies that they do not provide a lot of information about the mean velocity of the cluster. When a double-Gaussian fit is considered there is a significant difference. If the stars with large RV measurements are included, the best-fit second Gaussian has a contribution of \(4 \pm 2 \%\) with a mean velocity of \(3 \pm 20\) km s\(^{-1}\) and a velocity dispersion of \(32 \pm 43\) km s\(^{-1}\) (data not shown), so this second Gaussian now only provides a fit to outlying (probably bad) radial velocities, rather than suggesting a non-Gaussian velocity distribution of IC 348 like above. Adding this second Gaussian does not change the mean velocity or velocity dispersion of the first Gaussian in a significant way. Because of these outliers and the minor contribution of these stars with large RV measurements on the fit, we retain our cut in the measurement uncertainties for all future fits.

\begin{figure*}
\centering
\includegraphics[width=1\textwidth]{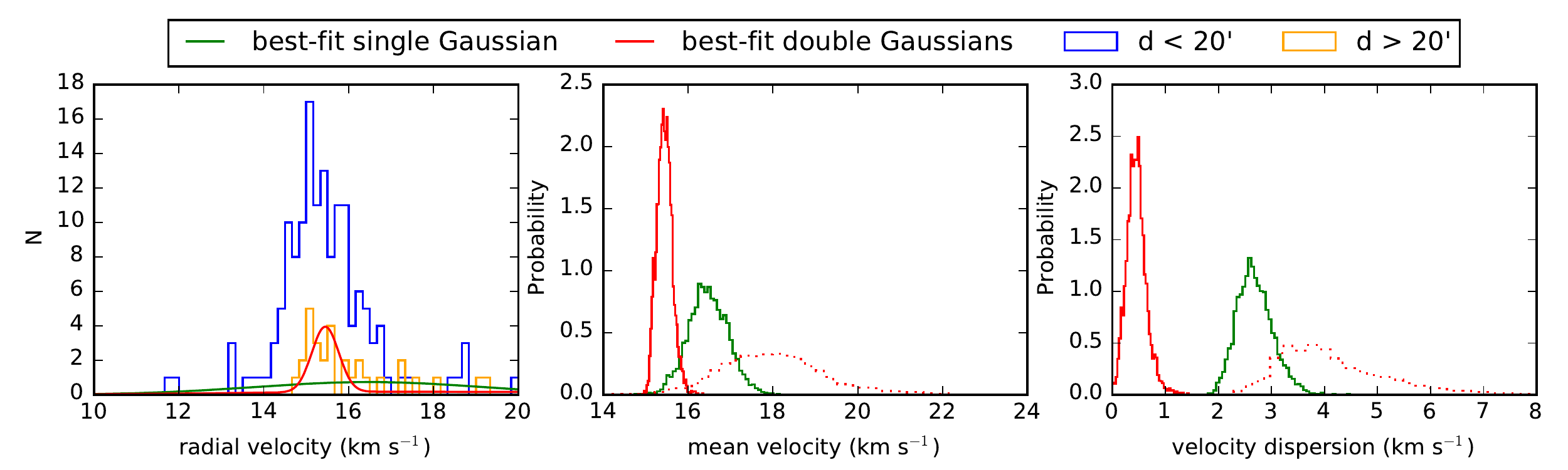}
\caption{\label{fig:outlier}Left panel: Histogram of all stars included in primary fit (i.e., within \(20\arcmin\) from cluster center; blue) and all stars with the same cuts, but beyond \(20\arcmin\) from the cluster center (orange). The best-fit single-Gaussian (green) and double-Gaussian (red) model to the stars beyond \(20\arcmin\) from the cluster center have been overplotted. Middle and right panel: Posterior probability distribution in the mean velocity and velocity dispersion of the single-Gaussian (green) and double-Gaussian model (red) for the stars beyond \(20\arcmin\) from the cluster center. For the double-Gaussian model the red lines show the posterior probability distributions of both Gaussians (solid and dashed).}
\end{figure*}

The results of fitting the RV distribution of the stars beyond \(20\arcmin\) (after making the same quality cuts) have been illustrated in Figure \ref{fig:outlier}. The RV distribution of these spatial outliers is poorly fitted by a single Gaussian (see green line in left panel) and hence requires a double-Gaussian fit. These Gaussians seem to have different mean velocities (Figure \ref{fig:outlier}), which suggests that they represent two different populations. One of these populations (\(41 \pm 11 \%\) of stars; \(\mu = 18.0 \pm 1.3\) km s\(^{-1}\); \(\sigma = 4.0 \pm 1.1\) km s\(^{-1}\)) appears to be inconsistent with either the velocity distribution of IC 348 or the much wider field velocity distribution (see e.g. slow rotators in upper panel of Figure \ref{fig:clean}). The mean velocity is consistent with the mean velocity of \(23.5 \pm 3.9\) km s\(^{-1}\) previously measured for the Perseus OB2 association \citep{Steenbrugge03}, which covers the observed region on the sky. The other population (\(59 \pm 11 \%\) of stars; \(\mu = 15.5 \pm 0.2\) km s\(^{-1}\); \(\sigma = 0.4 \pm 0.2\) km s\(^{-1}\)) is consistent with the RV distribution of IC 348, which suggests that either these stars are ejected from IC 348 or they formed outside of IC 348 from the same molecular cloud. If this extended population is also present in front and behind IC 348 along the line of sight, it might explain the non-Gaussianity observed in the radial velocity distribution in IC 348.

\subsection{Mass and virial state of IC 348}
\label{sec-4-3}
\label{sec:mass_virial}
\subsubsection{Stellar mass}
\label{sec-4-3-1}
\label{sec:star}
We need an estimate of the stellar and gas mass of IC 348 to determine its virial state. First we derive the stellar mass in IC 348. We will only consider stars more massive than 0.25 M\(_\odot\) (i.e., hotter than \(\sim 3300\) K), because these are typically brighter than H=13 for which our target catalogue of potential members seems to be complete (see completeness analysis in Section \ref{sec:target} and Figure \ref{fig:complete}).

First we derive the stellar mass of the stars with APOGEE spectra using the spectroscopic effective temperature and extinction-corrected absolute J-band magnitude from the Dartmouth isochrones. Here we include all stars targeted as potential members, except for the slow rotators (\(v \sin i < 5\) km s\(^{-1}\)), which are probably non-members as we argue in Section \ref{sec:samp}. For the stars with spectroscopic masses above 0.25 \(M_\odot\) the total mass adds up to about \(145 \pm 32\) M\(_{\odot}\) within 20", where the uncertainty comes from an assumed systematic uncertainty of 200 K on the effective temperature \citep{Cottaar14a}.

We then extend this mass estimate to the unobserved stars in the target catalogue. We assume that these unobserved stars have the same properties as the observed stars at the same \emph{apparent} H-band magnitude, because the priority in the targeting was based on this magnitude. For every unobserved star we select the 20 observed stars with the most similar apparent H-band magnitude and randomly assign one of their masses to the unobserved star. On average this leads to an additional mass of \(24 \pm 5\) M\(_\odot\) for cluster members with a mass above the 0.25 M\(_\odot\). Hence we estimate the total cluster mass for stars more massive than 0.25 M\(_\odot\) to be \(169 \pm 37\) M\(_\odot\).

\begin{figure}[htb]
\centering
\includegraphics[width=.9\linewidth]{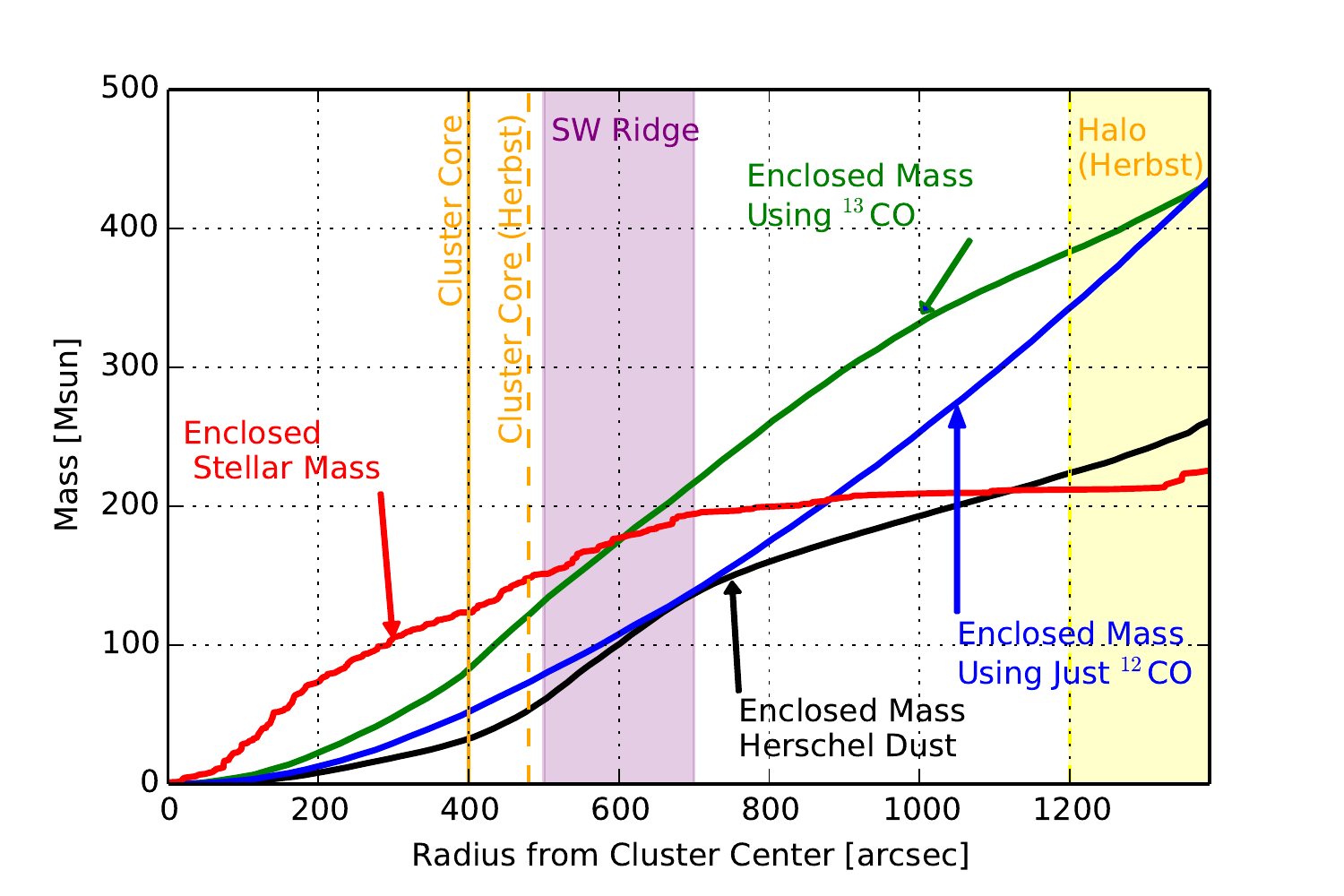}
\caption{\label{fig:mass}The enclosed stellar and gas mass in IC 348 as a function of distance from the cluster center. For the gas mass estimates from the Herschel dust emission maps, \(^{12}\)CO maps, and \(^{13}\)CO maps have been included. The core and halo radius of \citet{Herbst08} have been marked together with the half-mass radius adopted here of 6.3'. A dense gas clump called the southwestern ridge extends from about 500 to 700" to the south of IC 348.}
\end{figure}

Finally we estimate how much of the total cluster mass we missed by assigning a minimum stellar mass of 0.25 M\(_\odot\). For this we use the initial mass function (IMF) from \citet{Maschberger13}, which has been designed to match the classical IMF's from \citet{Kroupa01} and \citet{Chabrier03, Chabrier05}, but has the advantage of consisting of a single mathematically convenient continuous function. By integrating this IMF with the default parameters from \citet{Maschberger13} we find that stars with masses below 0.25 M\(_\odot\) contribute 17\% of the initial total mass. Correcting for this implies a final total stellar mass in IC 348 of \(204 \pm 45\) M\(_{\odot}\). The distribution of this mass has been plotted in Figure \ref{fig:mass} in red.

\subsubsection{Gas mass}
\label{sec-4-3-2}
\label{sec:gas}
We estimate the enclosed gas mass in IC 348 as a function of projected distance from the cluster center using three different methods. In the first, we use the publicly available Herschel data for Perseus from the Gould Belt Survey \citep{Andre10} which measures the thermal emission from dust at 100, 160, 250, 350 and 500 \micron. We convolve all these maps to the resolution of the coarsest map (500 \micron, 37") and for each pixel fit for the temperature and column density using a greybody fit, where the opacity/emissivity at each wavelength is taken from the \citet{Ossenkopf94} model for dust with thin ice mantles coagulating for 10\(^5\) years at a density of 10\(^{6}\) cm\(^{-3}\). This differs from the standard procedure of fitting submillimeter emission \citep[e.g.][]{Hildebrand83} because rather than assuming that emission from dust is a black-body modified by some power-law with exponent \(\beta\), it allows the dust emissivity to have the more complicated dependence on wavelength estimated by the \citet{Ossenkopf94} model. In addition, we do not assume that the emission is optically thin. We still assume a single temperature along the line of sight and that the dust optical properties are constant along the line of sight (and well described by the \citealt{Ossenkopf94} model), with all the concurrent uncertainty \citep[e.g.][]{Shetty09}. Following \citet{Hildebrand83}, we adopt a gas-to-dust mass ratio of 100.

We compare two alternate estimates of the gas mass in IC 348 using CO maps from the COMPLETE \citep{Ridge06} Survey. We use the relationships calibrated in \citet{Pineda08} to convert the CO emission to a total gas mass; these relationships are calibrated with reference to the COMPLETE near-infrared dust extinction maps, and thus ultimately also represent a dust mass with an assumed gas-to-dust mass ratio of 100. We use both the non-linear relationship to convert integrated intensity of \(^{12}\)CO (1-0) to gas mass (Eq. 19) and the linear relationship between the column density of \(^{13}\)CO (1-0) and gas mass (Eq. 21), where the latter calculation assumes that the excitation temperature of \(^{12}\)CO (1-0) and \(^{13}\)CO (1-0) are equal in order to estimate the optical depth of the \(^{13}\)CO (1-0) emission. These relationships are calibrated for different sub-regions in Perseus, and we use the parameters for IC 348 but note that the calibration in this region actually excludes the core of the cluster where the COMPLETE near-infrared extinction map is unreliable due to the large density of non-background stars. 

Figure \ref{fig:mass} shows these three different estimates for the enclosed mass, which show substantial disagreement among the methods, illustrating the fundamental uncertainty of this calculation. We adopt the Herschel dust mass primarily because the CO conversion relations are not well calibrated in the central region of the cluster where we most care about the enclosed gas mass. The presence of substantial protostellar outflows and shells in the vicinity of the IC 348 \citep{Arce10, Arce11} may also be complicating the calculation of gas mass from CO emission.
\subsubsection{Virial state}
\label{sec-4-3-3}
\label{sec:virial}
To determine whether IC 348 is expected to expand we compare the measured velocity dispersion of \(0.72 \pm 0.06\) km s\(^{-1}\) (or \(0.64 \pm 0.08\) km s\(^{-1}\) when fitting two Gaussians) to the velocity dispersion needed for virial equilibrium assuming spherical symmetry, which is given by:
\begin{equation}
\sigma_{\rm dyn} = \sqrt{\frac{G M}{\eta r_{\rm hm}}}, \label{eq:sdyn}
\end{equation}
where \(G\) is the gravitational constant, \(M\) is the total stellar and gas mass, \(r_{\rm hm}\) is the half-mass radius, and \(\eta\) is a structural parameter that depends on the density distribution. To estimate \(r_{\rm hm}\) and \(\eta\) we fit an \citet{Elson87} profile to the stellar distribution. This profile has a flat surface density distribution in the cluster core, which transitions to a power-law surface-density profile with slope \(-\gamma\) at a distance \(a\):
\begin{equation}
  \Sigma = \Sigma_0 \left(1 + \frac{R^2}{a^2}\right)^{-\gamma/2}, \label{eq:eff}
\end{equation}
where \(R\) is the projected distance from the cluster center and \(\Sigma_0\) is the central surface density. To minimize the extinction bias from the patchy molecular cloud on this fit we only include the stars with extinction-corrected J-band magnitude brighter than 12.5. Because we only have an extinction-corrected J-band measured for stars with APOGEE spectra, this limits ourselves to the subsample of observed member stars. This should be a minor effect, because the vast majority of these stars have an apparent H-band magnitude brighter than 12.5, which is a magnitude range over which we are mostly complete (Figure \ref{fig:complete}). We only fit the profile out to \(20\arcmin\) from the cluster center to minimize the effect of individual outliers. The density profile is fitted to the data by maximizing the likelihood (\(\mathcal{L}\)) of reproducing all stellar distances from the cluster centers:
\begin{equation}
  \mathcal{L} = A \prod_i R_i \Sigma(R_i), \label{eq:Lprof}
\end{equation}
where we multiply over all stars \(i\) at a distance \(R_i\) from the cluster center and the normalization of the probability distributions has been summarized by a single constant \(A\). This fit results in \(a = 4.1\)' and \(\gamma=3.1\), which corresponds to a half-mass radius of 6.3' or 0.47 pc at a distance to IC 348 of \(273 \pm 23\) pc \citep{Ripepi14}, where we assume that the distribution of stars with extinction-corrected J-band magnitude brighter than 12.5 is representative of the distribution of all stars. This matches the observed half-mass radius (Figure \ref{fig:mass}) reasonably well, which provides support to the adopted density profile. \citet{Portegies-Zwart10} showed that for \(\gamma \gtrsim 2.5\) the \citet{Elson87} profile consistently yielded \(\eta \simeq 10\), which we will adopt here.

Entering these values in equation \ref{eq:sdyn} we find a velocity dispersion of \(0.44 \pm 0.06\) km s\(^{-1}\) for virial equilibrium if we ignore the gas mass, while a cluster with a velocity dispersion \(\sqrt{2}\) higher (i.e. \(\sim 0.6\) km s\(^{-1}\)) would have a positive total energy, which could cause the cluster to unbind. Adding the gas mass to this estimate of the virial state is more tricky, because the gas is clearly not spherically distributed around the center of the stellar cluster with a large fraction contained in the south-western ridge (see Figure \ref{fig:target}), where its gravitational potential is less likely to contribute to holding IC 348 together, than if this gas was found in the center of IC 348. Projected onto the region of interest (within \(20\arcmin\) from the cluster center), there is a total of \(210 M_\odot\) of gas according to the Herschel dust maps, however only \(40 M_\odot\) of this gas is actually projected within one half-mass radius of the cluster center. Based on the spread in extinction observed for the cluster members in IC 348 we conclude that the stars are embedded within at least this gas. So as a lower estimate for the effect of the gas on the dynamical state of IC 348 we assume that this \(40 M_\odot\) has the same density profile as the stars for a total extra mass of \(80 M_\odot\), which leads to a velocity dispersion of \(0.51 \pm 0.04\) km s\(^{-1}\) with the cluster being unbound for a velocity dispersion greater than \(\sim 0.72\) km s\(^{-1}\). Finally as an extreme upper case we consider if all the gas of \(\sim 210 M_\odot\) had the same distribution as the stars (even though this is clearly inconsistent with the observed surface density distribution). In that case the velocity dispersion in virial equilibrium would be significantly higher at \(0.61 \pm 0.04\) km s\(^{-1}\) with the cluster being unbound at \(\sim 0.87\) km s\(^{-1}\). All these estimates of the velocity dispersion in virial equilibrium for different gravitational influences of the interstellar gas have been marked as vertical lines in the right panel of Figure \ref{fig:vdisp}.

\begin{table*}[htb]
\caption{\label{tab:energetics}Probability of various dynamical states for the single-Gaussian model (left) and double-Gaussian model (right) for various assumptions about the contribution to the gravitational potential of the gas.}
\centering
\begin{tabular}{|c|c|c|c|c|c|c|c|}
 & & \multicolumn{3}{c}{Single Gaussian} & \multicolumn{3}{c}{Double Gaussian}\\
Description & Energetics & no gas & low gas & high gas & no gas & low gas & high gas\\
\hline
Subvirial & \(E_{\rm kin} / E_{\rm pot} \leq 0.5\) & \(<\) 0.1\% & 0.2\% & 8\% & \(<\) 0.1\% & 3\% & 22\%\\
Supervirial, but bound & \(0.5 < E_{\rm kin} / E_{\rm pot} < 1\) & 19\% & 54\% & 88\% & 30\% & 64\% & 74\%\\
Unbound & \(E_{\rm kin} / E_{\rm pot} \geq 1\) & 81\% & 46\% & 5\% & 69\% & 34\% & 4\%\\
\end{tabular}
\end{table*}

The observed velocity dispersion was \(0.72 \pm 0.07\) km s\(^{-1}\) for the single-Gaussian model and \(0.64 \pm 0.08\) km s\(^{-1}\) for the double-Gaussian model. This is sufficiently high that the gas will have to significantly contribute to the gravitational potential for the cluster to be in virial equilibrium, with the observed velocity dispersion being more than 2-sigma higher than the velocity dispersion estimates with no gas or only a little gas (see Table \ref{tab:energetics}). This analysis suggests that IC 348 might be slightly supervirial, however the velocity dispersion does not appear to be so large that it can not regain virial equilibrium through expansion. 

\subsection{Velocity dispersion profile}
\label{sec-4-4}
\label{sec:radial}
In globular clusters the velocity dispersion is often found to drop with increasing distance from the cluster center \citep[e.g.][]{Watkins13} as expected from the Jean's equation. Such a trend would bias our velocity dispersion, because our completeness is lower in the cluster center, where the survey efficiency was limited by the collision radius of the input fibers. Here we explore how the velocity dispersion varies with distance from the cluster center under the assumption that this trend can be approximated by a second-order polynomial:
\begin{equation}
  \log(\sigma_c) = p_0 + p_1 d + p_2 d^2, \label{eq:rad_pol}
\end{equation}
where \(d\) is the distance from the cluster center and \(p_0\), \(p_1\), and \(p_2\) are the polynomial constants. We fit the polynomial to the logarithm of the dispersion to ensure that the dispersion remains positive. The mean velocity is still kept the same for all stars, which leads to a total of four parameters for the single-Gaussian model and 7 for the double-Gaussian model. Using this parameterization of the velocity dispersion and mean velocity we then maximize the likelihood of reproducing the observed velocity distribution (equation \ref{eq:conv}) and compute the uncertainties on these best-fit parameters (as well as their correlations) with an MCMC.

\begin{figure*}
\centering
\includegraphics[width=1\textwidth]{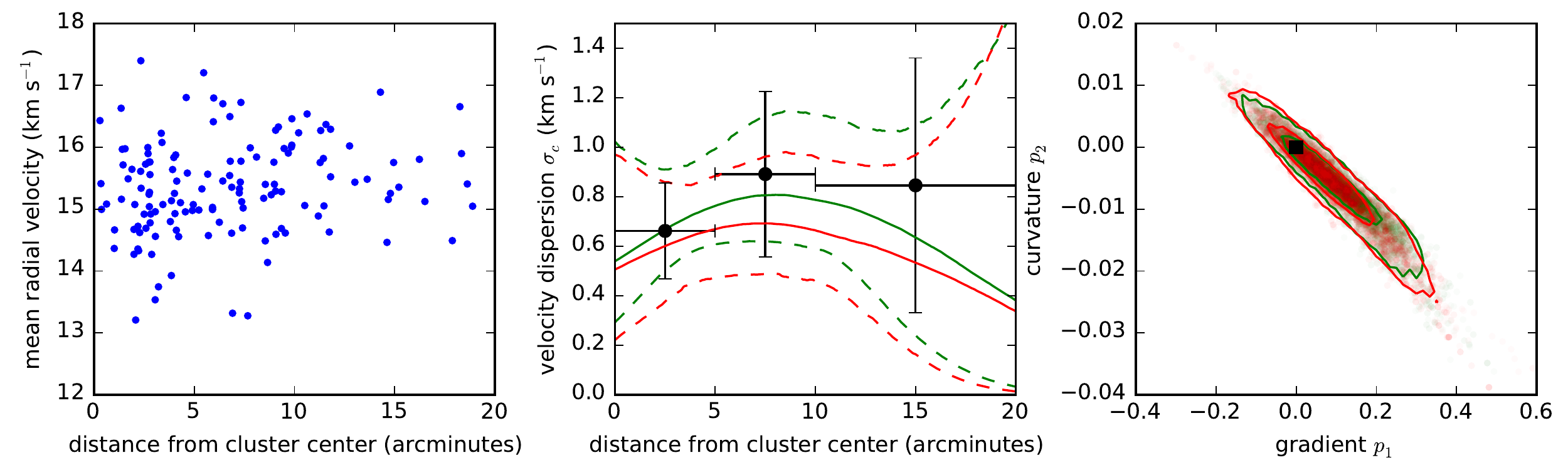}
\caption{\label{fig:rad_pol}(left) Mean radial velocity over all epochs versus distance from the cluster center for stars with no RV variability. (middle) Probability distribution of velocity dispersion with distance from the cluster center under the assumption that the true trend with velocity dispersion can be modeled by a second-order polynomial (equation \ref{eq:rad_pol}); the solid lines show the median velocity dispersion estimated as a function of distance, while the dashed lines show the 2-sigma uncertainties with the single-Gaussian model in green and the double-Gaussian model in red. The black dots show the best-fit velocity dispersion corrected for binary orbital motions for the stars in three radial bins (0-5', 5'-10', and 10'-20') with 2-sigma uncertainties. (right) Probability distribution of the gradient of the logarithm of the velocity dispersion with distance (\(p_1\)) and the curvature of this gradient (\(p_2\)); the transparent points show individual draws from the probability distribution from the MCMC with the 68\% and 95\% contours shown as solid lines. Both the single-Gaussian model (green) and the double-Gaussian model (red) are consistent within one sigma with a flat velocity dispersion profile (\(p_1 = p_2 = 0\) as marked by the black square).}
\end{figure*}

To visualize the result we compute the velocity dispersion profile for every point in the MCMC (ignoring the initial burn-in of the MCMC), so that we get the probability distribution of the velocity dispersion at any distance from the cluster center given the limitations of the second-order polynomial model. This is illustrated in the middle panel of Figure \ref{fig:rad_pol}. We find no evidence for variation of the velocity dispersion with projected distance from the cluster center, as illustrated by (1) that the probability distribution of the velocity dispersion remains flat within the uncertainties over the full distance range (red and green lines in center panel of Figure \ref{fig:rad_pol}), (2) that the velocity dispersion estimated in three radial bins are consistent (black dots in center panel of Figure \ref{fig:rad_pol}), (3) that both the gradient \(p_1\) and the curvature \(p_2\) are consistent with zero (right panel of Figure \ref{fig:rad_pol}), and (4) that the BIC has a higher value than for the global fit with a single velocity dispersion (see Table \ref{tab:BIC}), so that the extra parameters used to describe a polynomial trend of the velocity dispersion with distance does not improve the likelihood of the best fit sufficiently to be warranted.
\subsection{Mass segregation in velocity space}
\label{sec-4-5}
\label{sec:mass}
\citet{Muench07} and \citet{Schmeja08} showed that the massive stars in IC 348 are more centrally concentrated than the average cluster member. This implies that these stars are in closer orbits around the cluster center and might have a different velocity profile, although we found no evidence for such a dependence of the velocity dispersion profile on distance from the cluster center (Section \ref{sec:radial}). Similarly to our fit with distance from the cluster center we model the possible trend of velocity dispersion with stellar mass (\(M\)) with a polynomial:
\begin{equation}
  \log(\sigma_c) = p_0 + p_1 \log(M) + p_2 \log(M)^2. \label{eq:mass_pol}
\end{equation}
Here we fit both the logarithm of the velocity dispersion to ensure a positive velocity dispersion and the logarithm of the mass to stop the fit from being dominated by a few high-mass stars, which are less significant outliers in logarithmic space. 

\begin{figure*}
\centering
\includegraphics[width=1\textwidth]{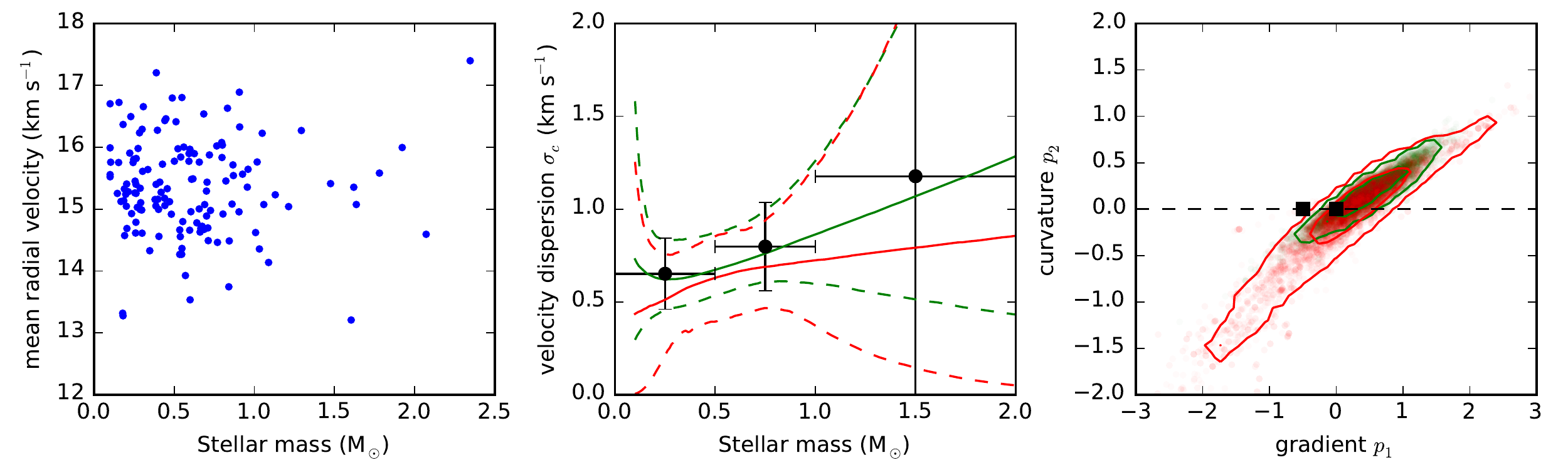}
\caption{\label{fig:mass_pol}Left panel: Trend of mean radial velocity over all epochs versus stellar mass. Middle panel: Probability distribution of the trend of velocity dispersion with mass estimated from the MCMC under the assumption of a second-order polynomial relation between the logarithm of the velocity dispersion and the logarithm of the mass (equation \ref{eq:mass_pol}); the colors illustrate whether a single-Gaussian model (green) or double-Gaussian model (red) was used with the mean dispersion marked by the solid line and the 2-sigma contours by the dashed lines. The black dots illustrate the velocity dispersion corrected for binary orbital motions estimated in three mass bins (\(M < 0.5 M_\odot\), \(0.5 M_\odot < M < M_\odot\), and \(M > M_\odot\)). Right panel: Probability distribution of the gradient \(p_1\) and curvature \(p_2\) of the trend of the logarithm of the velocity dispersion with the logarithm of the mass from the MCMC for the single-Gaussian model (green) and double-Gaussian model (red). The semi-transparent dots show the individual draws from the MCMC with the 68\% and 95\% contours shown as solid lines. The black squares mark a flat velocity dispersion profile (\(p_1 = p_2 = 0\)) and equipartition (\(p_1 = -0.5\) and \(p_2 = 0\)).}
\end{figure*}

The middle panel of Figure \ref{fig:mass_pol} illustrates the trend of the velocity dispersion with mass estimated from the MCMC computed under the constraint of the polynomial model of the velocity dispersion with mass (equation \ref{eq:mass_pol}). Although there is a suggestive increase of the velocity dispersion with mass this is not statistically significant with a flat velocity dispersion profile (\(p_1 = p_2 = 0\)) consistent with the data within one sigma (right panel of Figure \ref{fig:mass_pol}). We find that the velocity dispersion profiles diverge towards the edges of the covered mass range, which is likely caused by the tendency of polynomial fits to diverge at the extremes. This divergence prevents us from concluding anything solid about the dynamical state of the most massive stars (above \(1 M_\odot\)). Once again the BIC also suggests that these extra parameters do not significantly improve the fit to the velocity dispersion data (Table \ref{tab:BIC}).

If \(p_2\) is zero then equation \ref{eq:mass_pol} can be rewritten into a powerlaw:
\begin{equation}
  \sigma_c = e^{p_0} M^{p_1}.
\end{equation}
For \(p_2\) is zero we find that the slope of this powerlaw \(p_1\) is \(0.13 \pm 0.17\) for the single-Gaussian model (or \(0.28 \pm 0.45\) for the double-Gaussian model), which as noted above is fully consistent with flat (\(p_1=0\)). It is inconsistent at 2 sigma with all stars having the same kinetic energy (i.e., equipartition) in which case \(\sigma_c \propto M^{-1/2}\). Because dynamical evolution due to stellar interactions will drive a system towards rather than away from equipartition, this confirms the commonly held assumption that the initial velocity distribution of stars is better described by a flat velocity dispersion profile with mass than by equipartition.

If caused by dynamical evolution, mass segregation in young star clusters is expected to only affect the most massive stars, while the stars below a cluster-dependent threshold all have the same spatial distribution \citep{Allison10, Parker14}. The continuous polynomial model described above would provide a poor fit if the velocity dispersion was similarly constant over a broad range of masses with only the most massive stars having a different velocity dispersion. So we provide an additional fit of the velocity dispersion as a function of mass based on the simplest model that can capture such a discontinuity, namely the step function:
\begin{equation}
  \sigma_c = 
   \begin{cases}
      \sigma_{\rm low}  &\text{if } M < M_{\rm crit} \\
      \sigma_{\rm high} &\text{if } M \geq M_{\rm crit},
   \end{cases} \label{eq:step_mass}
\end{equation}
where \(\sigma_{\rm low}\) is the velocity dispersion below the mass cutoff \(M_{\rm crit}\) and \(\sigma_{\rm high}\) is the velocity dispersion above this cutoff.

\begin{figure}[htb]
\centering
\includegraphics[width=.9\linewidth]{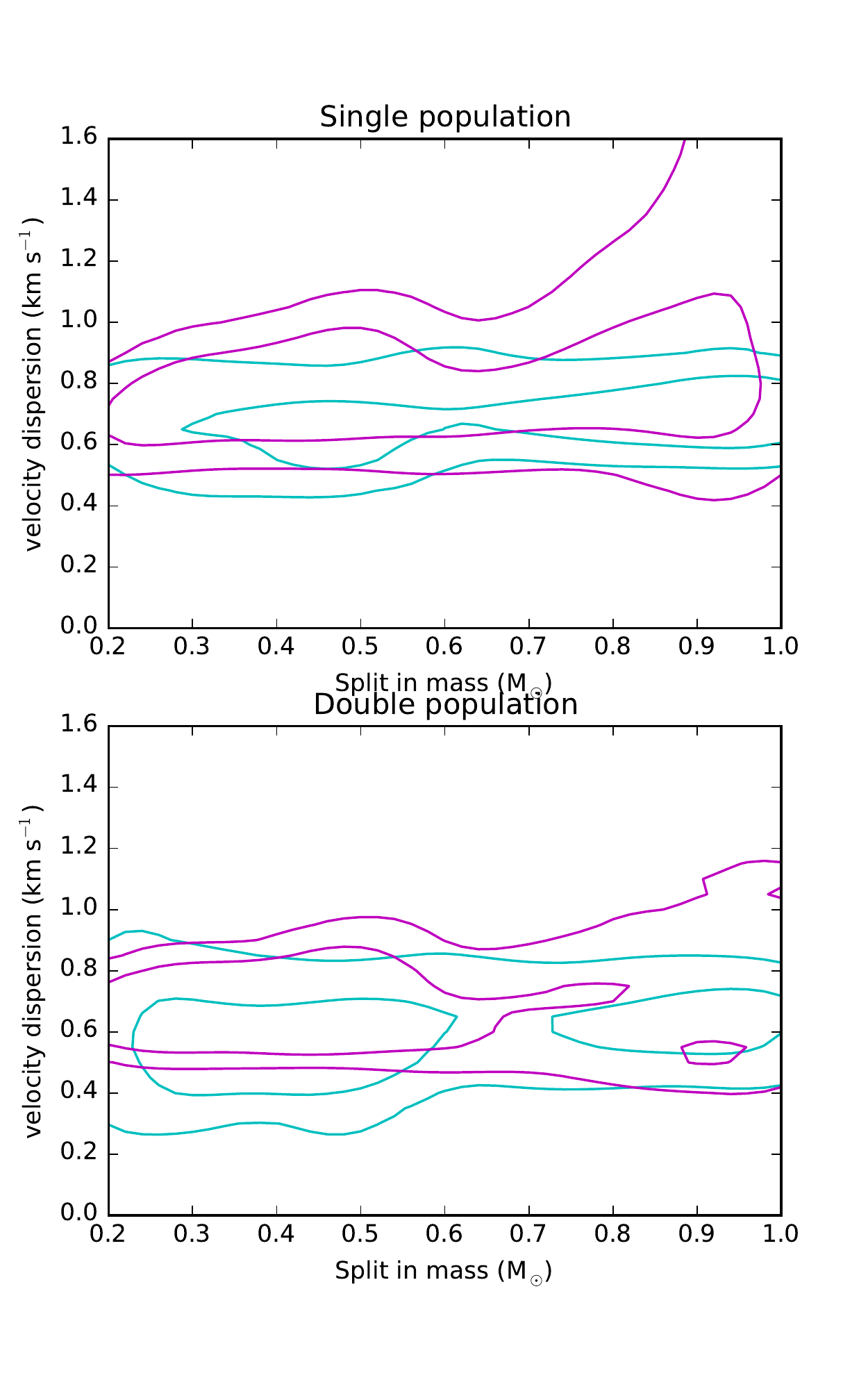}
\caption{\label{fig:mass_dep}Top panel: Posterior probability distribution of the velocity dispersion below the mass cut-off (\(\sigma_{\rm low}\); cyan) and above the mass cut-off (\(\sigma_{\rm high}\) magenta) plotted against the mass cut-off from the MCMC for the single-Gaussian model, assuming that the velocity dispersion can be described by a step function (equation \ref{eq:step_mass}. The lines show the 1- and 2-sigma uncertainties. Lower panel: Same as top panel, but for a model with two Gaussians.}
\end{figure}

Figure \ref{fig:mass_dep} shows the probability distribution from the MCMC of both the \(\sigma_{\rm low}\) and \(\sigma_{\rm high}\) plotted against the cut-off mass \(M_{\rm crit}\). At no cut-off mass \(M_{\rm crit}\) does the velocity dispersion for the lower-mass stars significantly differ from that of the higher-mass stars for either the single-Gaussian model (top panel) or double-Gaussian model (bottom panel), so even with this perhaps more appropriate model we do not find evidence for mass segregation in velocity space. Two local maxima to the likelihood function are found, one around a cutoff around \(0.4 M_\odot\) and one with a cutoff around \(0.8 M_\odot\). Neither of these significantly improves on the fit with a global velocity dispersion according to the BIC (Table \ref{tab:BIC}), which once again confirms that no mass segregation is found.

Despite the fact that we find no mass segregation, we note that the velocity dispersion is extremely poorly constrained in either model over the mass range over which we were most likely to see a segregation in velocity space (\(> 1 M_\odot\)). This poor constraint is partly because IC 348 only contains few massive stars, however it is exacerbated by the exclusion of the most massive stars from our analysis. Our analyzed subset included only 30 stars with an estimated mass above 1 \(M_\odot\). Future work that includes the velocities from all massive cluster members (either radial velocities from these or other spectra or proper motions) can further constrain this mass segregation. Because our sample contains so few of these stars, any mass segregation affecting only the velocities of these most massive stars will not significantly affect our observed velocity dispersion.
\section{Trends in mean velocity}
\label{sec-5}
\label{sec:mean}
\subsection{Velocity gradient with reddening}
\label{sec-5-1}
\label{sec:ext}
\begin{figure*}
\centering
\includegraphics[width=1\textwidth]{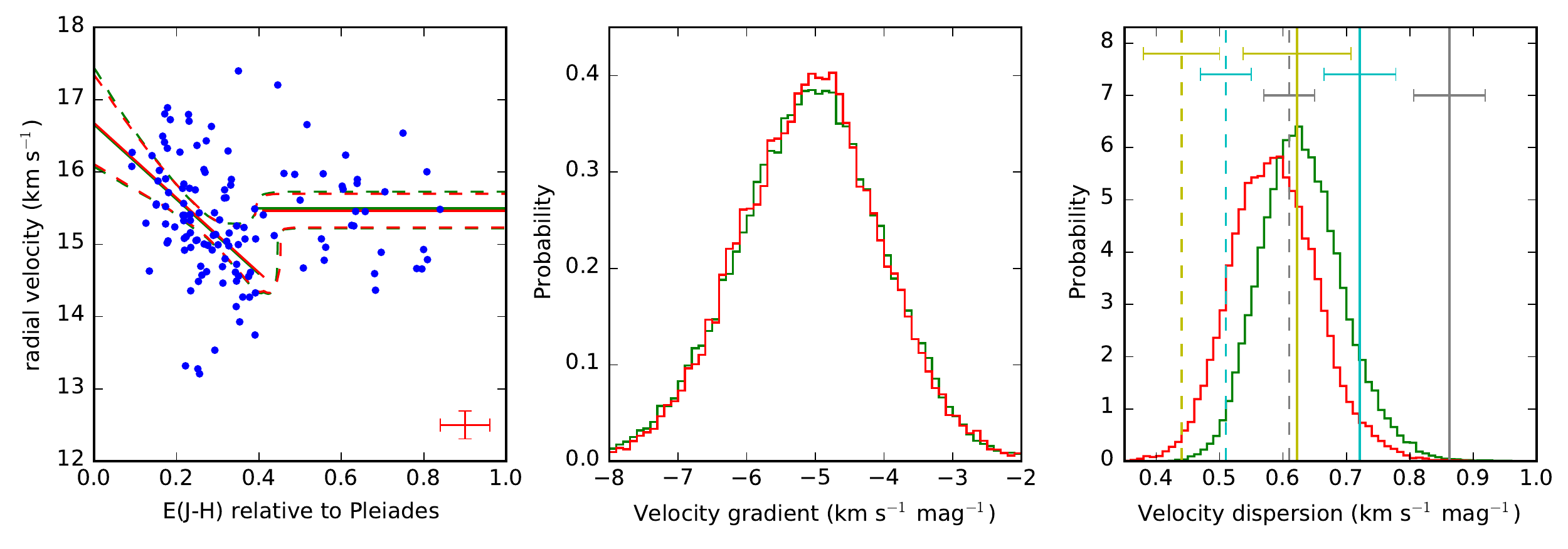}
\caption{\label{fig:EJH}(Left) Scatter plot of radial velocity versus stellar extinction with their typical uncertainties in the lower-right corner. The solid lines show the best-fit trend of velocity with extinction given our model of a velocity gradient of extinction for lowly extincted stars and a flat mean velocity above an extinction threshold. The dashed lines show the 95\% uncertainty on the mean velocity at that extinction from the MCMC. The green lines illustrate the single-Gaussian model, while the red lines illustrate the two-Gaussian model. The other panels show the probability distribution on the velocity gradient and velocity dispersion from the MCMC. The vertical lines in the right panel mark the velocity dispersion expected for virial equilibrium (dashed vertical line) and a dynamically unbound state with a net positive energy (towards the right of the solid vertical line) under various assumptions for the contribution of the gas mass to the dynamical state of IC 348, namely yellow for no contribution from the gas, cyan for a small contribution of \(80 M_\odot\), and gray for a large contribution of \(210 M_\odot\) (see Section \ref{sec:virial}).}
\end{figure*}

Ignoring for a moment local circumstellar material and the patchiness of the molecular cloud, the line-of-sight extinction measures how deeply a star is embedded within the gas in IC 348, and hence it gives us an indication of which stars are closer and further away from us along the line of sight. This would suggest that contraction or expansion of IC 348 along the line of sight might be directly detected through a correlation between the radial velocity and extinction of a star. To test this scenario we fit a linear relation between the mean velocity and the extinction up to an extinction-cutoff, above which the mean velocity is constant:
\begin{equation}
  \mu_c = 
  \begin{cases}
   \mu_{\rm low} + v_{\rm grad} \text{E(J-H)} &\text{if E(J-H)} < \text{E(J-H)}_{\rm crit} \\
   \mu_{\rm high}                             &\text{if E(J-H)} \geq \text{E(J-H)}_{\rm crit}
  \end{cases}  \label{eq:ext}
\end{equation}
This model provides an excellent fit to the data (left panel of Figure \ref{fig:EJH}) and is the only model providing a significantly better fit to the data than the basic global fit according to the BIC (see Table \ref{tab:BIC}). In both the single-Gaussian and double-Gaussian models a velocity gradient of \(-5.3 \pm 1.3\) km s\(^{-1}\) per magnitude in E(J-H) is detected up to an extinction of E(J-H) \(= 0.40 \pm 0.03\) mag (\(p < 10^{-4}\) that there is no gradient). The only difference between the single- and double-Gaussian models is again the velocity dispersion (Figure \ref{fig:EJH}). These velocity dispersions are lower than in the global fits with \(0.63 \pm 0.07\) km s\(^{-1}\) for the single-Gaussian model and \(0.55 \pm 0.07\) km s\(^{-1}\) for the double-Gaussian model, because some of the radial velocity variations are now modeled by the trend of the mean velocity with extinction. What this lowered velocity dispersion means for the dynamical state depends on the cause of the velocity gradient with extinction, which we shall return to in the discussion.

Without including an extinction cut-off in the model no velocity gradient is detected (\(0.0 \pm 0.3\) km s\(^{-1}\) mag\(^{-1}\)). The extinction cut-off of the velocity gradient with reddening can be understood by separating the reddening into a contribution from the molecular cloud and contributions from the stellar system. The stars in the background to IC 348 are estimated to have extinctions of \(E(J-H) \approx 0.4\) mag from both the COMPLETE extinction maps and the Herschel dust maps (see Figure \ref{fig:target}), so for stars with E(J-H) \(> 0.4\) mag we expect the reddening to be dominated by local contributions from the stellar system itself (i.e., emission from an inner proto-planetary disc or absorption from a disc or envelope). Below E(J-H) of 0.4 mag, however, the molecular cloud's contribution to the source's extinction is likely to be comparable to, and in most cases larger than, the source's local extinction.  As a result, reddening can be expected to correlate with position in the molecular cloud for stars with E(J-H) \(< 0.4\) mag, albeit with a non-negligible scatter due to varying contributions from local extinction.

So in this scenario the low-extinction stars (\(\text{E(J-H)} < 0.25\) mag) are preferentially located to the front of the molecular cloud with little local sources of reddening. Meanwhile the intermediate extinction stars (\(0.25 \text{ mag} < \text{E(J-H)} < 0.4\) mag) are located towards the back of the cloud with some contamination of stars in the front of the cloud with some local source of reddening. Finally the high-extinction stars (\(\text{E(J-H)} > 0.4\) mag) are distributed randomly throughout the cloud and have a significant local source of reddening. The blueshift of the low-extinction stars in the front of the molecular cloud compared with the intermediate-extinction stars towards the back of the cloud implies that the stars are moving towards each other and hence that the stars in IC 348 are converging along the line of sight.

The data presented so far can be explained by other scenarios as well. If the cluster is rapidly rotating and there exists a gradient in the reddening perpendicular to the rotation axis this could possibly cause the observed correlation between radial velocity and reddening. In Section \ref{sec:rot} we will see that the rotation in IC 348 is too small for this scenario. 

Alternatively the trend might not be statistically significant. Although the MCMC uncertainty on the velocity gradient (i.e., \(-5.3 \pm 1.3\) km s\(^{-1}\) mag\(^{-1}\)) imply a high statistical significance, these error bars were calculated under the assumption that the radial velocity of every star is an independent draw from the velocity distribution at its reddening. This assumption might not hold if the velocities are locally correlated. We look at that possibility in Section \ref{sec:corr}.

\subsection{Velocity gradient across the sky}
\label{sec-5-2}
\label{sec:rot}
\begin{figure}[htb]
\centering
\includegraphics[width=.9\linewidth]{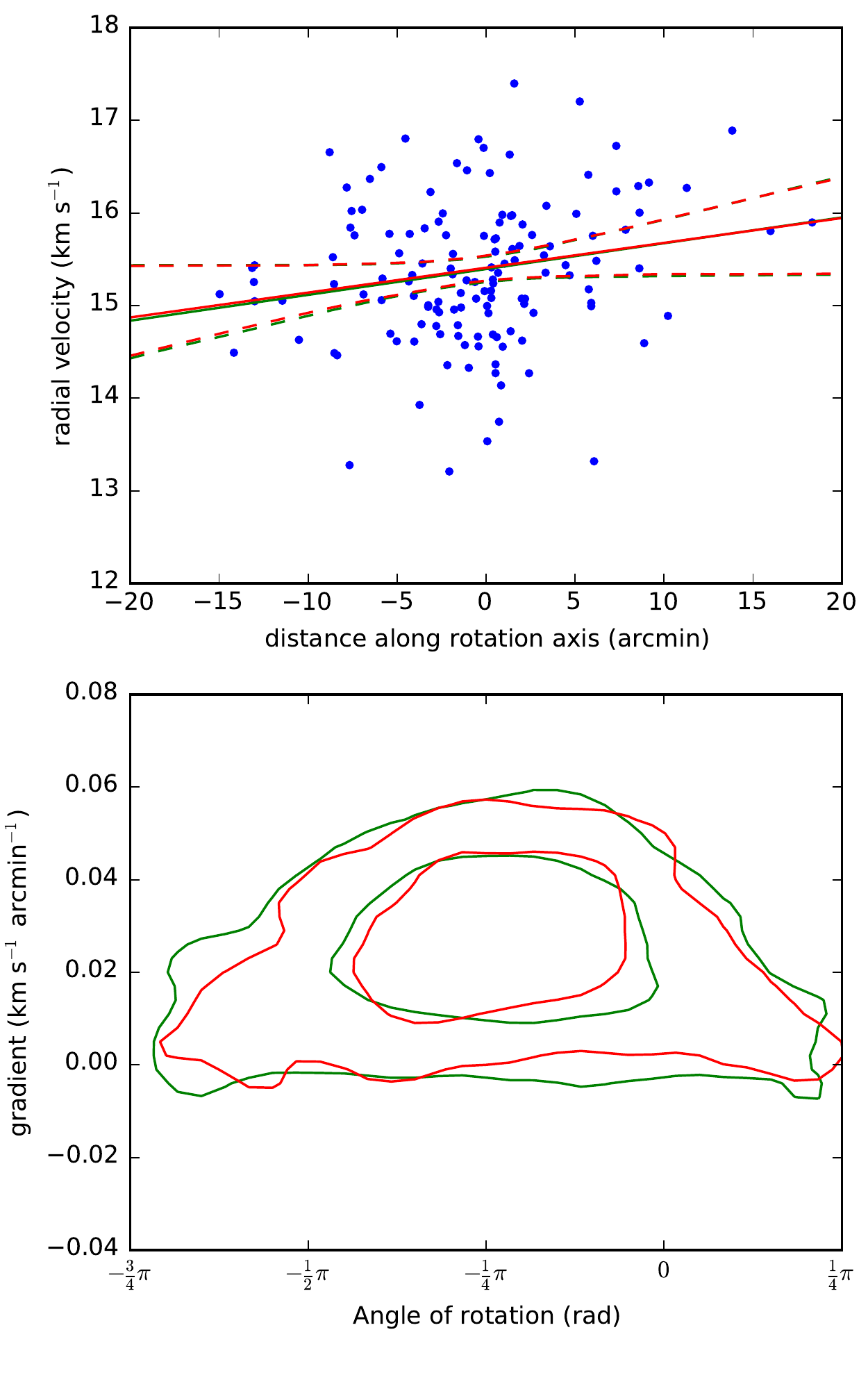}
\caption{\label{fig:rotation}The best-fit velocity gradient along a rotation axis from the southeast to the northwest with the 95\% limits shown as dashed lines (middle panel); and the probability distribution of the angle of rotation and the size of the velocity gradient (lower panel). The green lines illustrate the single-Gaussian model, while the red lines illustrate the double-Gaussian model.}
\end{figure}

A small velocity gradient is found with the mean velocity increasing from the southeast to the northwest (top panel of Figure \ref{fig:rotation}). To determine the significance of this trend, we add the possibility of solid-body rotation to our fit to the velocity distribution (see Section \ref{sec:method}) through the replacement of the global mean velocity with a different mean velocity for every star given by:
\begin{equation}
\mu_c = v_{\rm rot} r \cos(\theta - \alpha) + \mu, \label{eq:rot}
\end{equation}
where \(r\) and \(\theta\) are the positions for a star in circular coordinates around the cluster center (set at the median location of the analyzed stars: (RA, dec) = (56.13, 31.145)), \(v_{\rm rot}\) is the change in the mean velocity per arcminute, \(\alpha\) is the angle of the velocity gradient, and \(\mu\) is the global mean velocity.

The lower panel in Figure \ref{fig:rotation} illustrates the distribution of the MCMC, when fitting such a velocity gradient. Although the uncertainties on both the size (\(24 \pm 13\) m s\(^{-1}\) arcmin\(^{-1}\)) and the angle (\(-0.24 \pm 0.17 \pi\)) of the rotational gradient are large, in 94\% of the Markov chain there is a positive velocity gradient from the southeast to the northwest, suggesting at almost a 2-sigma level that there is a real gradient in this direction.

\subsection{Spatially correlated velocities}
\label{sec-5-3}
\label{sec:corr}
In the statistical analysis of the significance of the spatial trends of the velocity distribution discussed above, an implicit assumption was made that every star represented a random, independent draw of the velocity distribution. However, stars may form from a turbulent molecular cloud in which the velocities are strongly spatially correlated \citep{Larson81, Heyer04}. Here we check whether the stellar velocities in IC 348 are still spatially correlated, even though most stars will probably have completed several orbits through the cluster's potential well.

\begin{figure}[htb]
\centering
\includegraphics[width=.9\linewidth]{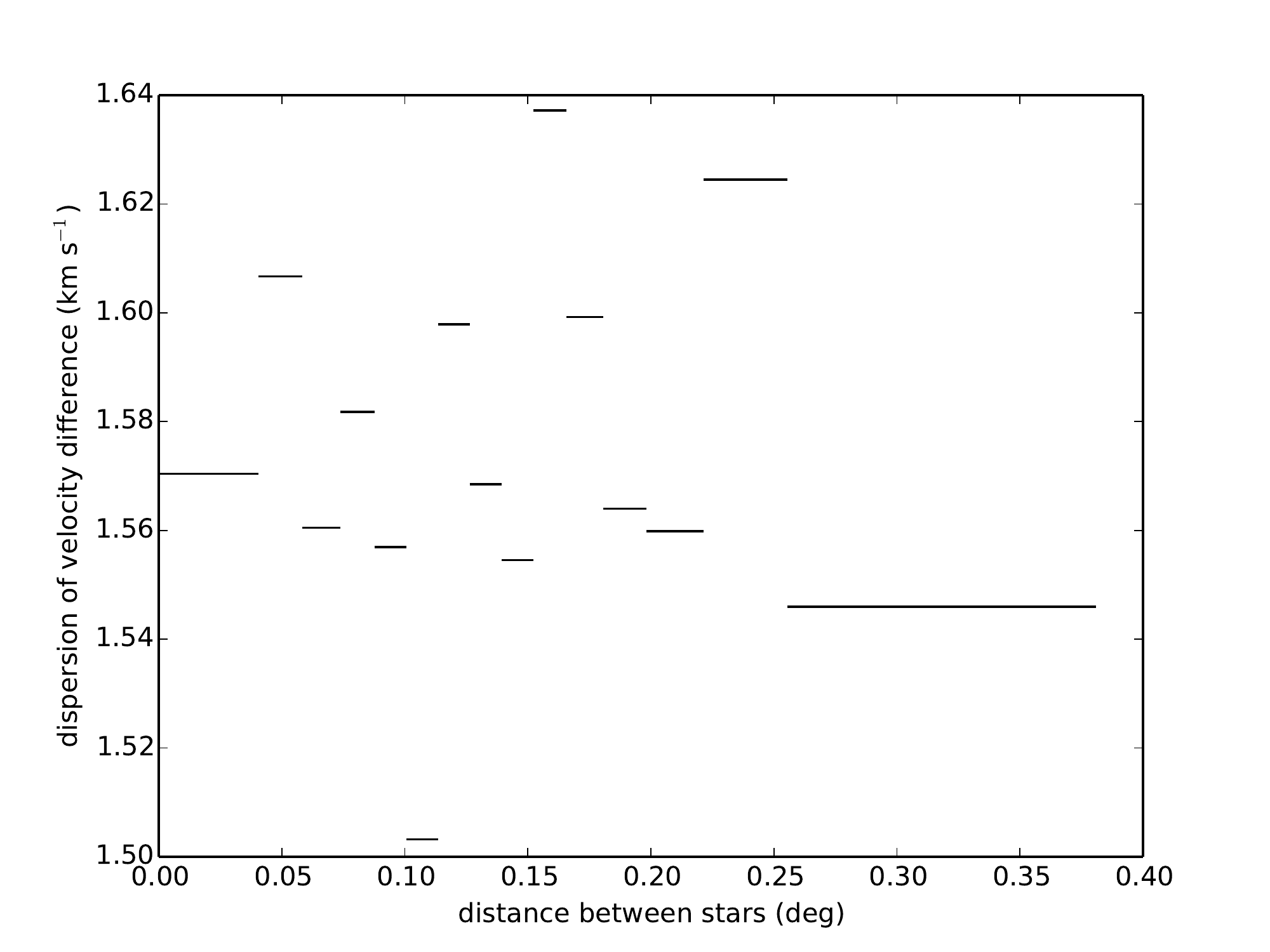}
\caption{\label{fig:vcorr}The dispersion of the velocity differences of stars with a projected separation in the range given by the length of the lines along the x-axis after cutting all velocity differences bigger than 4 km s\(^{-1}\) showing no obvious trend. Every bin contains about 3270 velocity differences between stars.}
\end{figure}

We test this correlation by computing the velocity differences between all stars and sorting them into equal-sized bins based on the projected distances between the stars. Figure \ref{fig:vcorr} shows for every bin the dispersion in velocity differences after cutting any velocity differences bigger than 4 km s\(^{-1}\) to minimize the effect of binary orbital motions. The dispersions are significantly higher than the global velocity dispersion (see section \ref{sec:global}), because the dispersion is not properly corrected for binary stars and the dispersion is computed for the velocity differences rather than the velocities, leading to an increase of \(\sqrt{2}\). Although the dispersion in molecular clouds increases by a factor of 4.5 as the distance scale goes up by a factor of 10 \citep{Heyer04}, we find that over the same increase in distance scale the stars show barely any increase in the velocity dispersion with an upper limit of \(\sim 10 \%\) suggested by the scatter. From this we conclude that there is no significant correlation left in the velocities between neighboring stars in IC 348. This may indicate that the cluster is dynamically old enough for the initial velocity substructure from the turbulent molecular cloud to have been erased.

\section{Discussion}
\label{sec-6}
\label{sec:disc}
\subsection{Scenarios for the velocity gradient with reddening}
\label{sec-6-1}
Here we explore possible causes of the RV trend with reddening found in Section \ref{sec:ext}. Our analysis indicates this correlation is statistically significant at the \(10^{-4}\) level; we welcome and encourage independent confirmation of this unexpected observational signature, which will ensure that the signature is astrophysical in nature and not the result of a statistical fluke or a systematic error in our observations or analysis. While we await this confirmation, we investigate the consequences and implications of this correlation, which is most likely caused by a correlation of both the RV and the reddening with a third potentially hidden parameter. The only scenario without such a third parameter is that there is a systematic offset of RV with reddening, which we explore in Section \ref{sec:syst}. The depth of a star along the line of sight is our main candidate for the third parameter that could correlate with both RV and reddening. Extinction naturally correlates with depth in the molecular cloud, if the cloud is sufficiently smooth. A correlation between depth and RV could be either explained by the convergence of two sub-clusters along the line of sight (Section \ref{sec:conv}) or a contraction of IC 348 along the line of sight (Section \ref{sec:contr}). Although both of these scenarios are consistent with the observations they require a large amount of fine-tuning as we will discuss below. This implies that one of these scenarios might explain the observations in IC 348, however if a similar convergence along the line of sight is found in more clusters in the future, new scenarios will have to be developed that explain the observations with less fine-tuning.

\subsubsection{Systematic offset in the velocities}
\label{sec-6-1-1}
\label{sec:syst}
One possible scenario that deserves exploring is that the trend is caused by a systematic offset in RV with reddening. The reddening causes stars to appear fainter, redder, and have relatively stronger diffuse interstellar bands. Here we argue there that any of these are unlikely to cause an offset in RV. 

A spectrum of a more reddened and hence fainter star has a larger contribution from telluric emission, which could potentially cause an offset in the RV. This brings to mind that the radial velocities of cooler (and hence fainter) stars were found to be redshifted by \citet{Cottaar14a}. However, as we argue in \citet{Cottaar14a} this redshift is unlikely to be caused by peculiarities in the data reduction, but rather by a systematic offset in the molecular (particular water) line lists for the coolest stars and furthermore the fainter stars in this case were found to be redshifted rather than blueshifted as would be needed to explain the trend with reddening seen in IC 348. The reddening also changes the slope of the observed spectra, however this slope is fitted out using a polynomial fit and is hence also unlikely to affect the RV measurement. Finally, although there are diffuse interstellar bands in the APOGEE spectral range \citep{Zasowski14}, they are weak and broad and hence unlikely to affect the measured RV.

A systematic offset with reddening could also not adequately explain, why the velocity gradient stops around E(J-H) = 0.4 mag with a constant mean velocity above. Finally a similar blueshift for more reddened stars is not found in the other clusters analyzed as part of the IN-SYNC ancillary program, which implies that this is a real trend specific to IC 348.
\subsubsection{Converging subclusters}
\label{sec-6-1-2}
\label{sec:conv}
Even if the trend of RV with reddening is caused by a trend of RV with depth in the molecular cloud, this does not automatically imply that the cluster as a whole is contracting. Alternatively IC 348 could consist of two sub-clusters, which overlap when projected along the line of sight. Such subclusters are often observed in star-forming regions and young clusters \citep[e.g.][]{Gutermuth08, Hacar13}. The subpopulation which is further away (with higher average reddening) could by chance be somewhat blueshifted. The lack of a gap in the E(J-H) distribution as well as the smooth gradient of radial velocity with E(J-H) suggest that there should be significant overlap in the reddening distribution of both sub-clusters. This scenario would require a very good alignment of the two sub-populations along the line of sight, so that they do not cause a larger velocity gradient with projected stellar location than is observed. Furthermore, such a scenario with multiple sub-populations seems unlikely given the smooth stellar distribution observed both spatially \citep{Schmeja08} and in velocity space (Section \ref{sec:corr}). Finally, in this scenario the two subpopulations could have had very different ages and other stellar parameters, which is not actually observed (see Section \ref{sec:contr}). Still we can not exclude this scenario based on the current data.
\subsubsection{Contraction of IC 348}
\label{sec-6-1-3}
\label{sec:contr}
Here we explore the scenario that IC 348 is contracting, which would explain the redshift of more strongly reddened stars, if they lie preferentially to the back of the molecular cloud. Like every molecular cloud the remaining gas in IC 348 is not smoothly distributed, but rather fairly patchy (Figure \ref{fig:target}; \citealt{Andre10}). So the trend of reddening with depth in the molecular cloud will differ per line of sight, which weakens the correlation between reddening and depth when averaged over the line-of-sights towards the different stars. This imperfect correlation would water down any real correlation between RV and depth, so that the strength of the observed correlation between RV and reddening is only a lower limit of the tighter correlation between RV and depth.

Still the observed correlation is already strong enough to put an interesting time constraint on the timescale of this contraction. The stars in the front of the cloud (with E(J-H) \(\approx 0.1\) mag) and in the back of the cloud (with E(J-H) \(\approx 0.35\) mag) are moving towards each other with \(1.3 \pm 0.4\) km s\(^{-1}\). If the cluster, which appears roughly circular, is actually spherical with a similar size of about 0.47 pc along the line of sight, the stars would pass each other in about 0.4 Myr, which puts an upper limit on how long this contraction can last. Although the size of the cluster and hence the contraction timescale could be much larger if the cluster is elongated (or in the two-population scenario described below), the larger the cluster is along the line of sight, the more unlikely it is that we do not detect this asymmetry in the projection on the plane of the sky.

This contraction timescale is much smaller than the age of the IC 348, so the cluster has only started contracting recently and is hence not caused by the initially subvirial state with which stars appear to form in molecular clouds \citep{Andre07, Kirk07}. It is very difficult to explain why contraction might initiate at a later stage. After all the early cluster evolution is characterized by mass loss, either stellar mass loss from stellar winds or supernovae or gas mass loss from the dispersing molecular cloud. Any such mass loss lowers the gravitational potential binding the cluster together, which should lead to expansion, not contraction \citep[e.g.][]{Hills80, Lada84, Goodwin06}.

One possible scenario to explain the contraction in IC 348 given this expected dominance of mass loss in determining the early cluster evolution is that IC 348 is revirializing. In this scenario a recent massive gas expulsion (e.g., supernova) has caused the cluster to become supervirial, but still bound. This results in the expansion of the cluster to a new virial equilibrium. Some N-body simulations \citep[e.g.][]{Goodwin06} suggest that this expansion might actually overshoot virial equilibrium, leading to a subvirial state and subsequent contraction back to virial equilibrium. Potentially we observed IC 348 just in a contraction stage of such an oscillation, however direct comparisons of the observed velocity trend with N-body simulations will be needed to test the viability of this scenario, which is beyond the scope of this paper. This scenario implies that IC 348 should have undergone a massive gas expulsion very recently (within \(\approx 1-2\) dynamical times or about 0.5-1 Myr), which should be observed in gas bubbles around IC 348 \citep[e.g.][]{Ridge06a, Arce10, Arce11}. Such a strong bubble is not apparent. Alternatively IC 348 could have been brought out of equilibrium in a recent merger of two or more subclusters.

One prediction of this scenario is that the redshifted, low-extinction stars should be drawn from the same population as the blueshifted, intermediate-extinction stars and hence should have the same stellar parameter distribution. To test this we split the stars with E(J-H) between 0 and 0.4 mag into two bins containing the same number of stars at \(\text{E(J-H)} = 0.25\) mag. We use KS-tests to find differences in the stellar parameters between these two populations. In Paper I we found that the surface gravity offset of a star from the median surface gravity at its effective temperature can be used to measure the relative sizes of stars at the same effective temperatures and hence should be sensitive to any age differences. The distribution of this relative surface gravity is consistent between the low-extinction and the intermediate-extinction populations (KS-test p-value of 11\% to be drawn by chance), which implies that we can not distinguish both populations based on age. Indeed we also find no difference in the \(R \sin i\) relative to the median \(R \sin i\) at the stellar effective temperature (KS-test p-value of 35\%) or in the \(v \sin i\) distribution between the two populations (KS-test p-value of 44\%). We do find that the stars with larger reddening are brighter after extinction-correction (KS-test p-value of 0.04\%), however this is probably caused by a bias in our sample selection and disappears if we only consider the stars with an extinction-corrected J-band magnitude brighter than 12.5. The stars with intermediate extinction do also appear to be slightly hotter (KS-test p-value of 2\%) and this trend remains at the same significance when we make a cut in extinction-corrected J-band magnitude of 12.5. However, it can be easily explained by stars for which we overestimate the temperature getting assigned somewhat higher extinctions and probably does not reflect a fundamental difference in the low-extinction and the intermediate-extinction populations. All of this implies that the low-extinction and intermediate-extinction stars are indeed consistent with being drawn from the same population.
\subsection{Virial state of IC 348}
\label{sec-6-2}
In Section \ref{sec:global} we found a velocity dispersion of \(0.72 \pm 0.06\) km s\(^{-1}\) (or \(0.64 \pm 0.08\) km s\(^{-1}\) if two Gaussians are fitted to the velocity distribution). We showed that there were only at most weak correlations of the velocity dispersion with distance from the cluster center or stellar mass, which implies that we can use these velocity dispersions as a representative value for the velocity dispersion of all stars in IC 348. These velocity dispersions seem to be higher than the velocity dispersion expected in virial equilibrium by about 2 \(\sigma\), unless the gas mass makes a much larger contribution to the velocity dispersion than expected (see right panel in Figure \ref{fig:vdisp} and Table \ref{tab:energetics}). This suggests that IC 348 is likely to be supervirial.

At first glance this supervirial state might appear inconsistent with the convergence of IC 348 along the line of sight. However, it does fit nicely in the revirialization scenario discussed in Section \ref{sec:contr}. If IC 348 is now contracting after an expansion overshot virial equilibrium, then the cluster will only be subvirial for the first half of this contraction. After that the contraction will pass virial equilibrium again, which leads to a contracting, supervirial cluster as observed.

In this analysis of the virial state we included both random motions and the systematic motions of RV with extinction in our estimate of the velocity dispersion, because we used the results of our fit with a single global mean velocity and velocity dispersion. In the scenario of a contracting cluster this is fine, because both the velocity gradient and the random motions represent contributions to the kinetic energy of IC 348 and hence should be included in the virial analysis \citep{Binney08}. This changes once we explain the RV gradient with extinction by converging subclusters.

If we have two converging subclusters, then the velocity gradient with extinction is an external motion and hence does not contribute to the virial state of either subcluster. To determine the virial state of these subclusters we hence have to use the velocity dispersion corrected for this RV gradient as measured in section \ref{sec:ext}. For the simple approximation of two sub-clusters, which each contain half of the stellar and gas mass observed in IC 348 and have the same velocity dispersion, we summarize the probability of being in various dynamical states in Table \ref{tab:corr_energetics}. In this case the velocity dispersion does appear more consistent with virial equilibrium. However, this should only be taken as a rough indication, not only because of the approximations about the properties of the two sub-clusters, but the velocity dispersion around the mean at a certain extinction measured here is likely to be a significant overestimate of the actual velocity dispersions of the sub-clusters, because at every extinction we probably have some contribution to the population from both sub-clusters.

\begin{table*}[htb]
\caption{\label{tab:corr_energetics}Probability of various dynamical states for a velocity dispersion corrected for the velocity gradient with extinction (see Table \ref{tab:energetics} for the uncorrected velocity dispersion).}
\centering
\begin{tabular}{|c|c|c|c|c|c|c|c|}
 & & \multicolumn{3}{c}{Single Gaussian} & \multicolumn{3}{c}{Double Gaussian}\\
Description & Energetics & no gas & low gas & high gas & no gas & low gas & high gas\\
\hline
Subvirial & \(E_{\rm kin} / E_{\rm pot} \leq 0.5\) & \(<\) 0.1\% & \(<\) 0.1\% & 0.1\% & \(<\) 0.1\% & 0.1\% & 4\%\\
Supervirial, but bound & \(0.5 < E_{\rm kin} / E_{\rm pot} < 1\) & 0.4\% & 5\% & 39\% & 6\% & 30\% & 73\%\\
Unbound & \(E_{\rm kin} / E_{\rm pot} \geq 1\) & 99.6\% & 95\% & 60\% & 94\% & 70\% & 24\%\\
\end{tabular}
\end{table*}
\subsection{Potential rotation}
\label{sec-6-3}
The final result we will discuss is the RV gradient observed across the plane of the sky from the northeast to the southwest. The interpretation of this velocity gradient is fundamentally different from that of the RV gradient with extinction, even though we also interpreted the latter as a correlation between radial velocity and stellar position. For the RV gradient in the plane of the sky the measured velocity is perpendicular to the projected stellar position, while in the RV gradient with extinction, the measured velocity is parallel to the measure of the stellar position. This means that only the latter is a direct measure of the contraction or expansion of the star cluster.

A common explanation for a RV gradient in the plane of the sky is rotation. Young clusters are thought to form from the merging of an initially substructured star-forming region. If these merging sub-clusters have significantly different mean velocities (compared to their velocity dispersion) the merged product will have significant angular momentum, so we do indeed expect rotation for such young clusters as IC 348 (and even much older open clusters if they are unable to get rid of their angular momentum in the subsequent dynamical evolution).

Another tempting scenario is that the RV gradient along the plane of the sky might actually be caused by the RV gradient with extinction. In the scenario of converging sub-clusters this gradient is easily explained by a slight offset between the two sub-clusters when projected on the plane of the sky. In the scenario of contraction in IC 348 the RV gradient across the plane of the sky could be explained if IC 348 is not perfectly spherical, but elongated along the velocity gradient. If the elongated tip of the stellar spatial distribution in the southwest is somewhat further away from us than in the northeast, than the net movement of stars towards the cluster center (i.e., convergence) will cause an average blueshift of stars in the southwest and redshift in the northeast as observed \citep{Proszkow09a}
\section{Conclusions}
\label{sec-7}
\label{sec:conc}
Using radial velocities determined from APOGEE spectra of 380 likely cluster members, we have measured the radial velocity distribution of the young (2-6 Myr) cluster IC 348. We find that two Gaussians provide a notably better fit to IC 348's velocity distribution than one Gaussian, even after explicitly including measurement uncertainties and the effect of binary orbital motions in the fits. The second component in the dual Gaussian fit reflects a non-Gaussian tail to the IC 348 velocity distribution, potentially caused by ejected stars or by young stars unrelated to IC 348 (i.e., from the distribution star-formation in the Perseus molecular cloud or from Perseus OB2). This dual Gaussian fit returns a velocity dispersion of 0.64 \(\pm\) 0.08 km s\(^{-1}\) for IC 348, moderately larger than expected if the cluster were in virial equilibrium (0.44-0.61 km s\(^{-1}\), depending on assumptions regarding the gravitational influence of gas in and around IC 348). Finally, we find in IC 348 a small velocity gradient of \(24 \pm 13\) m s\(^{-1}\) arcmin\(^{-1}\) across the plane of the sky.

Our analysis also identifies an intriguing kinematic signature of convergence in IC 348; if confirmed, this will be the first such detection in any star cluster. This convergence is inferred from a systematic blueshift for stars with intermediate reddening (in the back of the cloud) compared to stars with low reddening (in the front of the cloud). This kinematic trend does not extend to stars with large reddening values (i.e. \(\text{E(J-H)} > 0.4\) mag): these stars are more strongly reddened than the background stars in this field, implying that their observed colors are either dominated by highly localized (i.e circumstellar) extinction, or affected by infrared emission from hot gas, making their observed colors a poor proxy for depth within IC 348.

The combination of IC 348's super-virial velocity dispersion and kinematic convergence along the line of sight does not match theoretical predictions of the evolution of embedded star clusters. Any contraction due to subvirial initial conditions should have ended after a free-fall time (few 10$^5$ years), a timescale much smaller than IC 348's current age (2-3 or \(\sim\) 6 Myr). Afterwards, IC 348's evolution should be dominated by both stellar and gas mass loss and hence should be characterized by a super-virial expansion of the cluster. We speculate that the super-virial convergence we detect in IC 348 could be caused by a recent expulsion of gas from the cluster. Such an event could cause IC 348 to expand rapidly, overshoot the state of virial equilibrium, and then begin to contract and overshoot virial equilibrium again, at which point we would have a contracting, supervirial cluster as we observe for IC 348.

An alternative scenario for the correlation we see between RV and reddening is that the structure we identify as IC 348 is, in reality, a chance alignment of two subpopulations of stars moving towards each other along the line of sight (Section \ref{sec:conv}). This scenario seems inconsistent with the lack of spatial and velocity substructure observed in IC 348, however, and also requires an unusual agreement between the stellar parameters (i.e., age and rotational properties) of the two subpopulations. Proper motions measured by Gaia should be able to distinguish between these scenarios, as the converging sub-populations model suggests no contraction in the plane of the sky, while the revirialization scenario does imply a similar contraction in the plane of the sky. Both of these scenarios require a specific temporal or spatial alignment to explain IC 348's super-virial and converging state, however, motivating the need to find alternative scenarios for the presented observations.

\section{acknowledgements}
\label{sec-8}
MC derived the stellar parameters, analyzed the observed velocity distribution as discussed here and wrote the manuscript. KRC, JCT and MRM conceived the program's scientific motivation and scope, led the initial ancillary science proposal, oversaw the project's progress and contributed to the interpretation of the findings discussed herein; KRC also led the target selection and sample design process. DLN assisted in the interpretation of the APOGEE data products and reduction algorithms, particularly those related to radial velocity measurements. JBF assisted with the sample selection for the dynamical analysis and the gas mass estimate. NDR selected the targets in Orion A. KMF assisted with target selection. SDC and GZ oversaw the design of the APOGEE plates utilized for IN-SYNC observations. SRM and MFS contributed to defining the scope and implementation plan for this project, and with JCW developed and provided high level leadership for the broader APOGEE infrastructure that enabled this science. Furthermore we thank Rory Smith, Stella Offner, and Richard J. Parker for helpful comments and suggestions and thank Gus Muench and Luisa Rebull for help in the target selection of respectively IC 348 and NGC 1333.

MC and MRM acknowledge support from the Swiss National Science Foundation (SNF). Funding for SDSS-III has been provided by the Alfred P. Sloan Foundation, the Participating Institutions, the National Science Foundation, and the U.S. Department of Energy Office of Science. The SDSS-III web site is \url{http://www.sdss3.org/}.

SDSS-III is managed by the Astrophysical Research Consortium for the Participating Institutions of the SDSS-III Collaboration including the University of Arizona, the Brazilian Participation Group, Brookhaven National Laboratory, Carnegie Mellon University, University of Florida, the French Participation Group, the German Participation Group, Harvard University, the Instituto de Astrofisica de Canarias, the Michigan State/Notre Dame/JINA Participation Group, Johns Hopkins University, Lawrence Berkeley National Laboratory, Max Planck Institute for Astrophysics, Max Planck Institute for Extraterrestrial Physics, New Mexico State University, New York University, Ohio State University, Pennsylvania State University, University of Portsmouth, Princeton University, the Spanish Participation Group, University of Tokyo, University of Utah, Vanderbilt University, University of Virginia, University of Washington, and Yale University.
\bibliographystyle{aa}
\bibliography{global}
\end{document}